\documentclass[12pts,
aps,
floatfix,
letterpaper,
prx,
longbibliography,
singlecolumn,
reprint,
superscriptaddress]{revtex4-2}
\usepackage[up,bf,raggedright]{titlesec}

\usepackage[sectionbib]{bibunits}

\usepackage{newtxtext,newtxmath}

\usepackage[colorlinks=true, linkcolor=blue]{hyperref}

\usepackage{lipsum}
\usepackage{graphicx} 
\usepackage{amsfonts}
\usepackage{amsmath}
\usepackage{braket}
\usepackage{gensymb}
\usepackage{hyperref}
\usepackage{dsfont}
\usepackage[capitalize]{cleveref}

\usepackage[utf8]{inputenc}

\date{\today}

\begin{document}

\title[2D transmons with lifetimes and coherence times exceeding 1 millisecond]{2D transmons with lifetimes and coherence times exceeding 1 millisecond}

\author{Matthew P. Bland}
\altaffiliation{These authors contributed equally to this work.}
\affiliation{Princeton University, Department of Electrical and Computer Engineering, Princeton, NJ 08544, USA}
\author{Faranak Bahrami}
\altaffiliation{These authors contributed equally to this work.}
\affiliation{Princeton University, Department of Electrical and Computer Engineering, Princeton, NJ 08544, USA}
\author{Jeronimo G. C. Martinez}
\affiliation{Princeton University, Department of Electrical and Computer Engineering, Princeton, NJ 08544, USA}
\author{Paal H. Prestegaard}
\affiliation{Princeton University, Department of Electrical and Computer Engineering, Princeton, NJ 08544, USA}
\author{Basil M. Smitham}
\affiliation{Princeton University, Department of Electrical and Computer Engineering, Princeton, NJ 08544, USA}
\author{Atharv Joshi}
\affiliation{Princeton University, Department of Electrical and Computer Engineering, Princeton, NJ 08544, USA}
\author{Elizabeth Hedrick}
\affiliation{Princeton University, Department of Electrical and Computer Engineering, Princeton, NJ 08544, USA}
\author{Alex Pakpour-Tabrizi}
\affiliation{Princeton University, Department of Electrical and Computer Engineering, Princeton, NJ 08544, USA}
\author{Shashwat Kumar}
\affiliation{Princeton University, Department of Electrical and Computer Engineering, Princeton, NJ 08544, USA}
\author{Apoorv Jindal}
\affiliation{Princeton University, Department of Electrical and Computer Engineering, Princeton, NJ 08544, USA}
\author{Ray D. Chang}
\affiliation{Princeton University, Department of Electrical and Computer Engineering, Princeton, NJ 08544, USA}
\author{Ambrose Yang}
\affiliation{Princeton University, Department of Electrical and Computer Engineering, Princeton, NJ 08544, USA}
\author{Guangming Cheng}
\affiliation{Princeton University, Princeton Materials Institute, Princeton, NJ 08544, USA}
\author{Nan Yao}
\affiliation{Princeton University, Princeton Materials Institute, Princeton, NJ 08544, USA}
\author{Robert J. Cava}
\affiliation{Princeton University, Department of Chemistry, Princeton, NJ 08544, USA}
\author{Nathalie P. de Leon}\email{npdeleon@princeton.edu}
\affiliation{Princeton University, Department of Electrical and Computer Engineering, Princeton, NJ 08544, USA}
\author{Andrew A. Houck}\email{aahouck@princeton.edu}
\affiliation{Princeton University, Department of Electrical and Computer Engineering, Princeton, NJ 08544, USA}

\maketitle
\onecolumngrid
\textbf{Materials improvements are a powerful approach to reducing loss and decoherence in superconducting qubits because such improvements can be readily translated to large scale processors. Recent work improved transmon coherence by utilizing tantalum (Ta) as a base layer and sapphire as a substrate \cite{place2021new}. The losses in these devices are dominated by two-level systems (TLSs) with comparable contributions from both the surface and bulk dielectrics~\cite{crowley2023disentagling}, indicating that both must be tackled to achieve major improvements in the state of the art. Here we show that replacing the substrate with high-resistivity silicon (Si) dramatically decreases the bulk substrate loss, enabling 2D transmons with time-averaged quality factors ($Q$) exceeding $\mathbf{1.5 \times 10^{7}}$, reaching a maximum $Q$ of $\mathbf{2.5 \times 10^{7}}$, corresponding to a lifetime ($T_{1}$) of up to 1.68~ms. This low loss allows us to observe decoherence effects related to the Josephson junction, and we use improved, low-contamination junction deposition to achieve Hahn echo coherence times ($T_{\mathbf{2E}}$) exceeding $T_{1}$. We achieve these material improvements without any modifications to the qubit architecture, allowing us to readily incorporate standard quantum control gates. We demonstrate single qubit gates with 99.994\% fidelity. The Ta-on-Si platform comprises a simple material stack that can potentially be fabricated at wafer scale, and therefore can be readily translated to large-scale quantum processors.}

Recent advances in the performance and scale of superconducting qubit processors have enabled early demonstrations of quantum error correction~\cite {Acharya2025, krinner2022realizing, Gong2021Experimental, sivak2023realtime, ofek2016extending}, quantum simulation~\cite{barends2015digital, kandala2017hardware, marcos2013superconducting}, and quantum many-body physics~\cite{zhang2022synthesizing, mi2022time, kollar2019hyperbolic, andersen2024thermalization}. In some of the most advanced quantum error correction experiments with superconducting qubits~\cite{Acharya2025}, roughly half of the measured logical error budget is related to single qubit coherence; reducing the physical error rates by a factor of two would improve the logical performance of a $d = 27$ code by a factor of 10$^4$. Therefore, improvements in qubit coherence that preserve the basic underlying transmon architecture will yield major advances in processor performance. Most efforts aimed at achieving millisecond coherence times have focused on avoiding loss and decoherence through novel qubit designs, such as fluxonium qubits~\cite{FLuxoniumQubitHighQ}, Kerr-cat qubits~\cite{KerrCat}, and zero-pi qubits~\cite{0piqubit}, which require radically new processor architectures and gate schemes. Recent progress in superconducting qubit materials and fabrication has enabled the realization of 2D transmon qubits with quality factors exceeding 1.0 $\times 10^{7}$~\cite{place2021new, crowley2023disentagling, ganjam2024surpassing, wang2022towards, sivak2023realtime, gao2024establishing,gordon2022environmental, Deng2023titanium, sivak2023realtime, Biznárová2024, Bal2024, Kono2024, tuokkola2024methods}. These advances have been driven by a renewed interest in improving fabrication, removing material contamination, and choosing materials such as Ta that are robust to post-fabrication processing to enable device cleaning.

It has been recently shown that state-of-the-art Ta superconducting resonators at low microwave powers and low temperatures exhibit losses that are dominated by two-level systems (TLSs), and that these TLS losses arise from both the surfaces of the device and the bulk substrate with comparable contribution \cite{crowley2023disentagling}. In particular, the sapphire substrate exhibits a dielectric loss tangent $\tan \delta$ = $(1.3\pm 0.2) \times 10^{-7}$, which limits the $Q$ of 2D transmons to 7.7 $\times 10^{6}$, in line with recent demonstrations of Ta qubits on sapphire~\cite{place2021new, crowley2023disentagling, ganjam2024surpassing, wang2022towards, sivak2023realtime}. A natural strategy for improving qubits is to seek alternative substrates with lower dielectric loss, such as different grades of sapphire~\cite{dielectricDipper} or high resistivity silicon~\cite{Zhang2024accepter, Lozano2024LowLoss, martinis2014ucsb}.

Here we demonstrate methods for growing $\alpha$-phase Ta films on high-resistivity Si, and we fabricate superconducting resonators to measure the TLS-related loss contributions of the surfaces and bulk substrate. We find that although surface losses are comparable to Ta-on-sapphire resonators, there is no apparent saturation in $Q$ with increasing size, and therefore the bulk loss is notably lower. This reduced bulk loss translates to improved transmon qubit performance. We demonstrate a qubit with a $T_{1}$ up to 1.68 $\pm$ 0.1~ms, corresponding to a quality factor $Q = \omega \times T_1 = 2.5 \times 10^{7}$. Across an 88 hour period, this qubit has a time-averaged $T_{1}$ (denoted $T_1^{\text{avg}}$) of 1.0~ms and a time-averaged $Q$ (denoted $Q^{\text{avg}}$) of $1.5 \times 10^7$. This material system is exceptionally robust: across 45 qubits,  the mean $T_1^{\text{avg}}$ is $0.45 \pm 0.18$~ms and the mean $Q^{\text{avg}}$ is $9.7 \times 10^{6}$. These improved qubit lifetimes reveal decoherence pathways related to the aluminum/aluminum oxide (Al/AlO$_x$) Josephson junction, and we observe that $T_{\textrm{2E}}<T_{1}$ for many qubits. By reducing hydrocarbon contamination while depositing and oxidizing the Al/AlO$_x$ junction in ultrahigh vacuum (UHV) conditions, the qubit coherence time dramatically improves, and we achieve  $T_{2E}>T_{1}$ for most qubits. Using dynamical decoupling, we deduce that shot noise related to spurious photons in the readout resonator is a significant source of the remaining noise in qubits with UHV-deposited junctions. Because both improvements are related to the underlying materials and device fabrication, they can be readily translated to 2D transmon gate operations, and we demonstrate single qubit gate infidelities of (6.4$\pm$0.3)$\times10^{-5}$.

\begin{figure}
   \centering
  \includegraphics[width=0.7\textwidth]{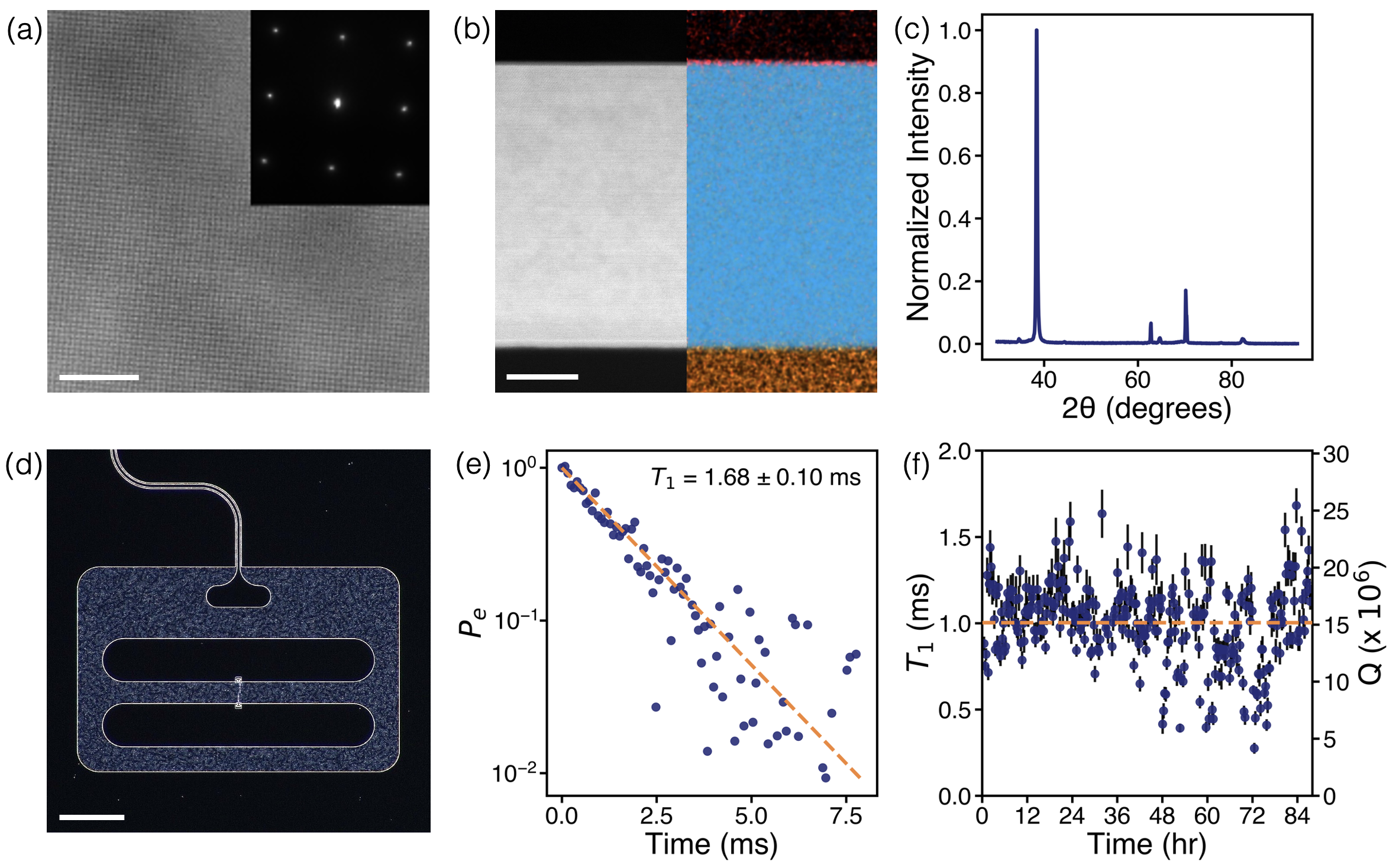}
  \caption{\textbf{Ta-on-Si Transmon Qubits with Relaxation Times Exceeding 1 ms}:
  (a) Cross-sectional scanning transmission electron microscope (STEM) image of Ta film deposited on Si substrate. Local area diffraction pattern (inset) confirms the BCC structure of $\alpha$-Ta with growth oriented in the $<$110$>$ direction. Scale bar represents 5~nm.
  (b) High-angle annular dark-field (HAADF) STEM (left) and false-colored energy-dispersive x-ray spectroscopy (EDS) elemental mapping (right) images of Ta film on Si, showing presence of Si (orange), Ta (blue), and oxygen (red). The EDS results confirms absence of any oxide at Ta-Si interface. Scale bar represents 50~nm.
  (c) X-ray diffraction (XRD) pattern for an $\alpha$-Ta film deposited on Si. The peaks at 38.4$^{\circ}$ and 82.3$^{\circ}$ correspond to $<$110$>$ and $<$220$>$ Ta peaks, respectively, and the peak at 70.1$^{\circ}$ corresponds to $<$400$>$ Si. The XRD pattern does not show evidence of $\beta$-phase Ta or silicides. The additional peaks at 44.4$^{\circ}$ and 64.6$^{\circ}$ are contamination peaks from the x-ray source. 
  (d) Optical microscope image of a transmon qubit with Ta capacitor pads and ground plane (black) on a silicon substrate (dark gray) and Al/AlO$_x$ Josephson junction (white). Scale bar represents 200 $\mu$m. 
  (e) Qubit population as a function of delay time showing a maximum lifetime $T_{1}^{\text{max}}$ = 1.68 $\pm$ 0.1~ms, corresponding to a quality factor of $Q^\text{max} = 2.5 \times 10^{7}$. 
  (f) Fitted $T_{1}$ for individual measurements recorded over time. Over an 88 hour period, the average lifetime $T^\text{avg}_{1}$ of 1.00 $\pm$ 0.01 ms, corresponding to an average quality factor of $Q^{\text{avg}} = 1.5 \times 10^{7}$.
  }

   \label{fig1:T1}
\end{figure}

Depositing high-quality $\alpha$-Ta films on Si requires careful control over the deposition conditions with particular attention to the cleanliness of the heteroepitaxial interface~\cite{Lozano2024LowLoss}. We first remove the native SiO$_{2}$ layer from the silicon substrate with dilute HF, and then we immediately transfer the substrate into the sputtering chamber to avoid oxide regrowth. The background pressure of the UHV sputtering chamber is around 2$\times 10^{-9}$ Torr. After loading the sample from the loadlock, the pressure increases to 5$\times 10^{-9}$ Torr, and we sputter a thin layer of Ta on the chamber walls as a gettering step to reduce the base pressure. The Ta films are then deposited at temperatures of 600 - 650 $^{\circ}$C, which results in $\alpha$-Ta with an out-of-plane $<$110$>$ orientation (Fig.~\ref{fig1:T1}(a, c)), and we verify that there is no oxide layer at the Ta-Si interface (Fig.~\ref{fig1:T1}(b)). The superconducting transition temperature measured by DC resistivity is 4.2~K, as expected for $\alpha$-Ta (Extended Data Fig.~\ref{Ta-characterization}). When we deposit Ta films with higher background base pressure, we observe minority formation of the $\beta$ phase or tantalum silicide. We use photolithography and argon/chlorine reactive ion etching to pattern the Ta films for transmon qubits (Fig.~\ref{fig1:T1}(d)) and resonators. For qubits, Al/AlO$_{x}$ Josephson junction fabrication proceeds via electron-beam lithography followed by double-angled Al deposition in an electron-beam evaporator (see Methods).

Ta qubits fabricated on high resistivity ($>$20 k$\Omega$-cm) Si exhibit remarkably long lifetimes. Our highest measured $T_{1}$ is 1.68 $\pm$ 0.10 ms which corresponds to $Q = 2.5 \times 10^{7}$~(Fig.~\ref{fig1:T1}(e, f)). We then perform repeated $T_{1}$ measurements over time to capture time-varying device performance. Over 88~hours, this qubit has a time-averaged lifetime $T_{1}^{\textrm{avg}} = 1.00 \pm 0.01$~ms, corresponding to a time-averaged quality factor $Q^{\text{avg}} = 1.5 \times 10^{7}$~(Fig.~\ref{fig1:T1}(f)). This material stack and device fabrication process is robust and reproducible. We measure 45 qubits over 9 separate chips for over three days, and for each qubit, we determine its maximum quality factor over time $Q^{\text{max}}$ and time-averaged quality factor $Q^{\text{avg}}$ (Tab.~\ref{table:T1s}). Across all measured qubits, the mean value of $Q^{\textrm{avg}}$ is 9.74$\times 10^6$ and the mean value of $Q^{\textrm{max}}$ is 1.60$\times 10^7$. Fluctuations in $T_1$ over time are commonly observed in transmon qubits~\cite{McRae2021reproducible, Klimov2018fluctuations}. To quantify these fluctuations, we define the \textit{span} of $T_1$ time fluctuations for a given qubit to be the difference between the qubit's first and third quartile $T_1$ over time, normalized by the qubit's time-averaged $T_1$. Across all 45 measured qubits, the mean span is $36\%$. 

\begin{figure}
   \centering
  \includegraphics[width=0.45\textwidth]{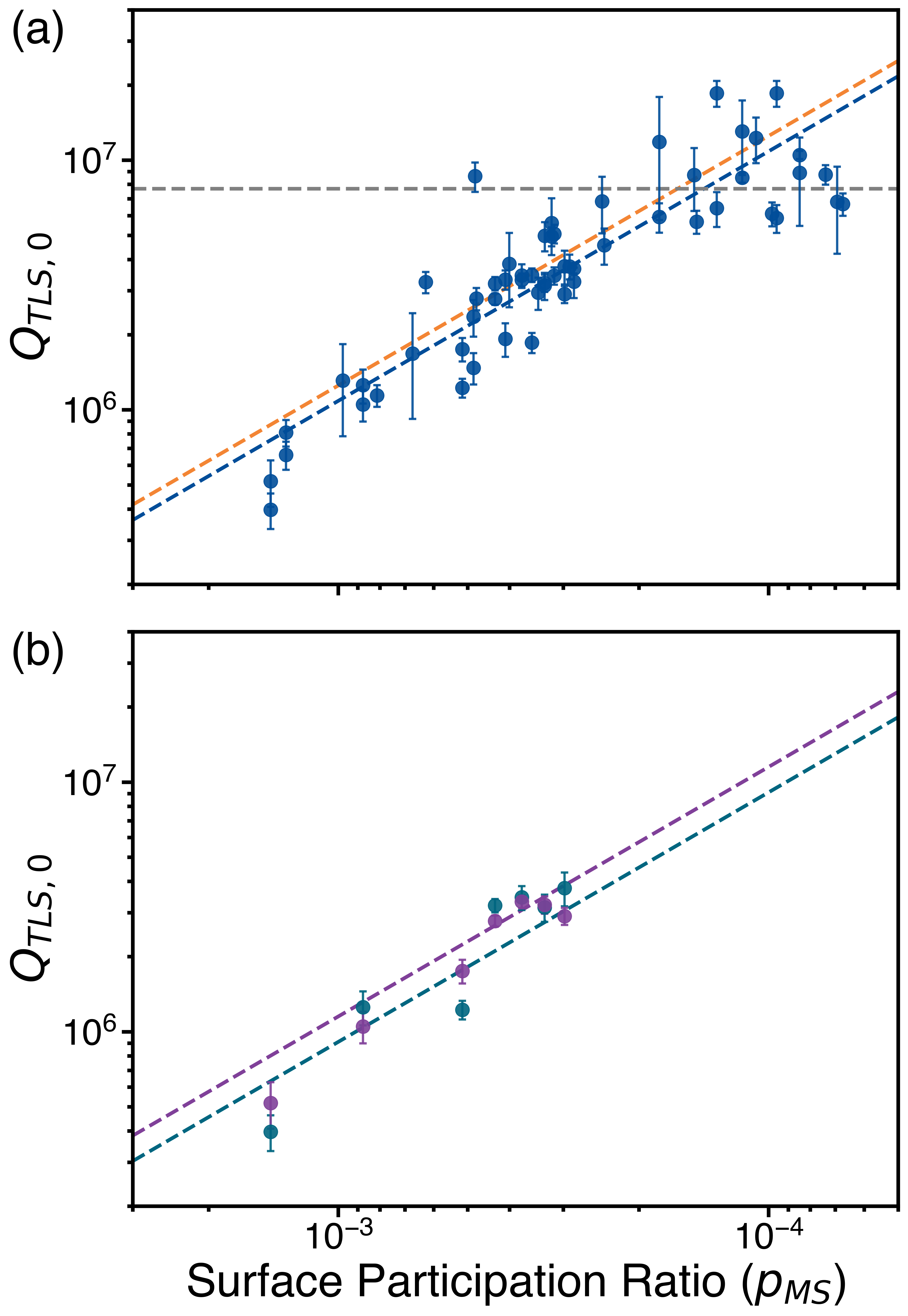}
  \caption{\textbf{Disentangling sources of loss}:
(a) $Q_{\text{TLS,}0}$ (blue circles) as a function of the metal-substrate surface participation ratio ($p_{MS}$). Surface participation ratios scale similarly with capacitor pitch, so we choose $p_{MS}$ for consistency with prior literature~\cite{wang2015spr,crowley2023disentagling}. Dashed lines represent fitted surface (blue) loss tangents for Ta-on-Si resonators. For comparison, surface (orange) and bulk (gray) loss tangents are plotted from previous work on state-of-the-art Ta-on-sapphire resonators ~\cite{crowley2023disentagling}. The surface losses are similar for resonators fabricated on the two substrates; however, for the larger devices (smaller SPRs) on high-resistivity Si, we see multiple instances of $Q_{\text{TLS,}0}$ that are higher than the bulk limit of sapphire.
(b) $Q_{\text{TLS,}0}$ as a function of $p_{MS}$ for Ta-on-Si resonators with estimated SiO$_{x}$ thickness of 0.3~nm (teal) and 1.5~nm (purple). Both data sets come from the same device, before and after exposing the device to ambient conditions for 65 days. The estimated surface loss for 0.3~nm (teal dashed line) and 1.5~nm (purple dashed line) are (11.0$\pm$1.8)$\times 10^{-4}$ and (8.7$\pm$0.6)$\times 10^{-4}$, respectively.}
   \label{fig2:wangplot}
\end{figure}

To investigate the sources of loss in our Ta-on-Si qubits, we measure resonators to isolate the contributions of the material system without the Josephson junction. We measure the temperature- and power-dependent behavior of Ta-on-Si resonators with varying surface participation ratio (SPR)~\cite{wang2015spr}, and we fit the data to a model that quantifies the separate contributions of TLSs, quasiparticles, and other loss channels~\cite{crowley2023disentagling}. We observe that in the millikelvin temperature and single-photon power regime, the dominant source of loss is TLSs  (see supplementary material). We therefore focus on a fit parameter that is inversely related to the linear absorption due to TLSs, $Q_{\textrm{TLS,0}}$ (Fig.~\ref{fig2:wangplot}(a)). (We note that $Q_\textrm{TLS,0}$ can slightly underestimate the total $Q$ for qubits because TLSs can saturate at low microwave powers, below the average operating power of qubits \cite{crowley2023disentagling}.) As the SPR decreases, $Q_\textrm{TLS,0}$ increases, consistent with the TLS loss being located in a thin layer at a surface or interface. The extracted surface loss tangent is similar to what was previously observed for Ta-on-sapphire resonators~\cite{crowley2023disentagling}. For the largest devices (lowest SPRs), $Q_\textrm{TLS,0}$ continues to increase with the same scaling, in contrast to observations of Ta-on-sapphire resonators, and we do not observe a saturation in $Q$ due to dielectric loss in the bulk substrate. Taking the standard deviation of $Q_\textrm{TLS,0}$ around the fitted surface loss tangent, we estimate a lower bound on the bulk-limited $Q_\textrm{TLS,0}$ of 5.3$\times 10^7$. We therefore conclude that the increased lifetime of Ta-on-Si qubits is enabled by the low bulk dielectric loss of high-resistivity Si.

Another potential source of loss in this material system is the formation of amorphous silicon oxide (SiO$_x$) at the exposed silicon surface under ambient conditions. To quantify the contribution of the surface oxide to loss, we compare the SPR dependence of $Q_\textrm{TLS,0}$ for Ta-on-Si coplanar waveguide resonators with 0.3~nm SiO$_x$ to resonators with 1.5~nm SiO$_x$ (Fig.~\ref{fig2:wangplot}(b)). The latter oxide thickness represents the observed oxide growth after 65 days in ambient conditions (Extended Data Fig.~\ref{SI:SIOx}). The estimated surface loss tangents for 0.3~nm and 1.5~nm are (11.0$\pm$1.8)$\times 10^{-4}$ and (8.7$\pm$0.6)$\times 10^{-4}$, respectively. The estimated surface losses are within each other's statistical uncertainties, from which we infer that SiO$_{x}$ is a minor contributor to surface loss. Based on the fit error of the surface loss tangent, the oxide loss due to a 1.5~nm thick SiO$_{x}$ layer is less than 10\% of the total surface loss.

\begin{figure}
   \centering
  \includegraphics[width=0.8\textwidth]{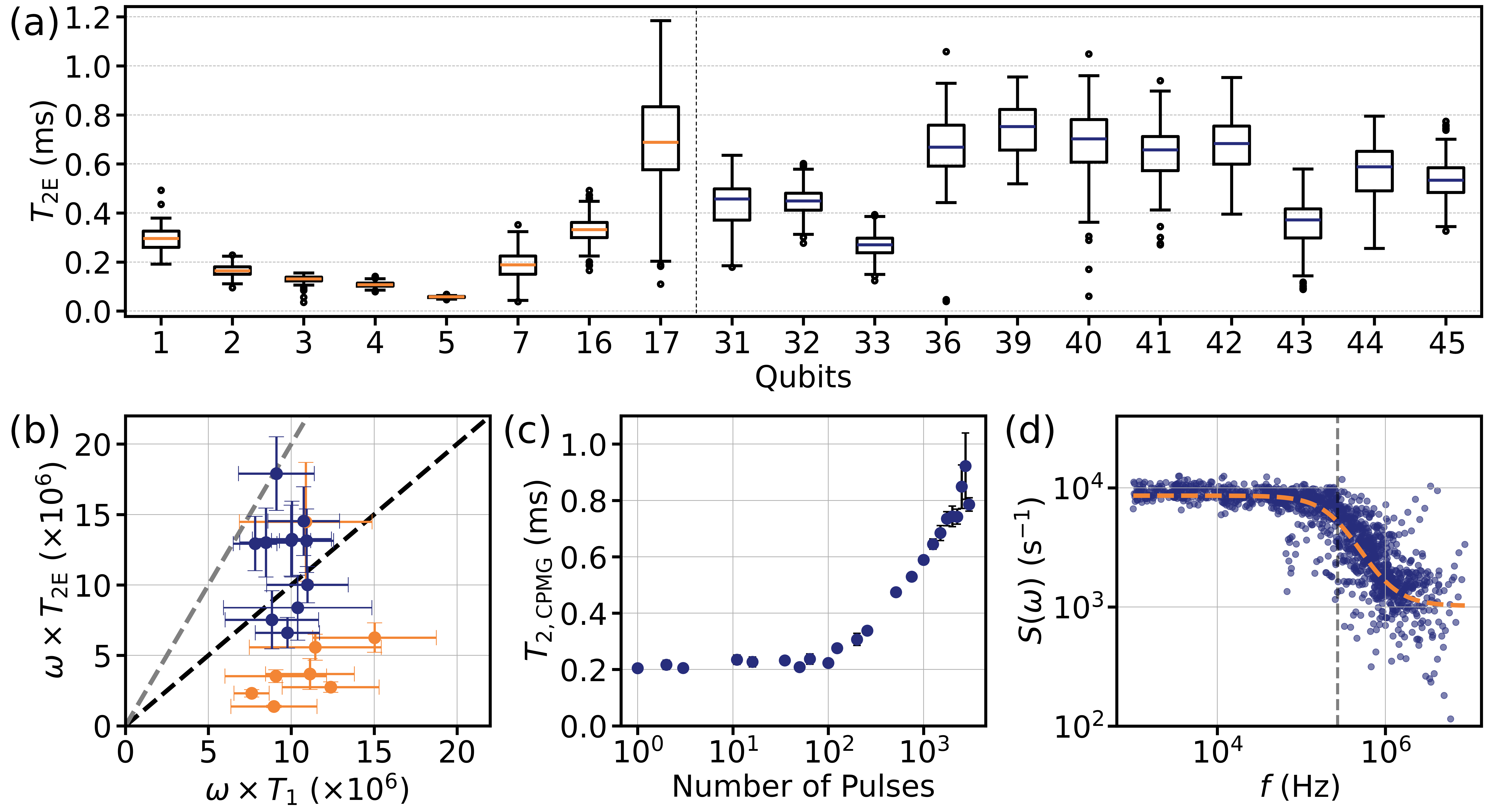}
  \caption{\textbf{Improving Qubit Coherence}: 
  (a) Time-varying statistics of Hahn echo coherence time $T_\text{2E}$ for 19 qubits across 6 fabricated chips. The numbers on the x-axis correspond to the same qubits as in Table~\ref{table:T1s}. Box and whisker plots depict median value (line), lower and upper quartiles (box), extrema (whiskers), and outliers (open circles) of $T_{2 \text{E}}$ over several days of recording. After optimizing junction fabrication in a UHV deposition chamber, the reliability of fabricating devices with long $T_\text{2E}$ improved dramatically (devices to the right of the gray dashed line).
  (b) Scatter plot of $T_\text{2E}$ versus $T_{1}$ multiplied by the qubit resonance frequency for individual qubits with junctions deposited in HV (orange) and UHV (purple) chamber conditions. The black and gray dashed lines refer to $T_\text{2E} = T_{1}$ and $T_\text{2E} = 2T_{1}$, respectively. The average $T_\text{2E}$ for junctions deposited in the HV chamber is $(0.45 \pm 0.37)T_{1}$ compared to $(1.2 \pm 0.4)T_{1}$ for junctions deposited in the UHV chamber.
  (c) Coherence time ($T_{2,\text{CPMG}}$) as a function of the number of CPMG dynamical decoupling pulses. For a small number of $\pi$-pulses, the $T_{2,\text{CPMG}}$ remains constant at a value of 0.2~ms. Beyond 100 pulses, $T_{2,\text{CPMG}}$ increases with $\pi$-pulse number and approaches the 2$T_{1}$ limit at 3000 pulses.
  (d) Qubit noise spectral density extracted from CPMG data. The flat portion of the spectral density at low frequencies corresponds to the constant $T_{2,\text{CPMG}}$ values for $<$100 $\pi$-pulses. Beyond 299 kHz, the noise spectral density decreases, with the cutoff closely matching the resonator decay rate of $\kappa/2\pi=288$~kHz (vertical dashed gray line). The orange dashed line is a fit based on a model of dephasing from coupling to thermal photons in the resonator.
  }

   \label{fig3:CPMG}
\end{figure}

The low dielectric loss of the Ta-on-high-resistivity Si platform enables more sensitive measurements of the contributions of the Al/AlO$_{x}$ Josephson junction to decoherence. We measure the Hahn echo coherence time ($T_\textrm{2E}$) and obtain a time-averaged Hahn echo coherence time $T_\textrm{2E}^{\text{avg}}$ over 3 days of recording for each measured qubit. Across all measured qubits, $T_\text{2E}^{\text{avg}} = (0.45 \pm 0.37) T_{1}$, with one outlier qubit achieving $T_\textrm{2E}^{\text{avg}}$ = 1.3 $T_1$ (Fig.~\ref{fig3:CPMG}(a) orange lines). Furthermore, $T_\textrm{2E}$ fluctuates in time (Fig.~\ref{fig3:CPMG}(a) orange lines, box and whisker plot); across all devices, the mean span of $T_{2 \text{E}}$ time fluctuations is $23\%$ (where span is defined analogous to $T_1$ fluctuations). The large $T_\textrm{2E}$ time fluctuations likely arise from strongly coupled TLSs in the junction, which experience spectral diffusion into the qubit resonance frequency \cite{Schlor2019TLSDecoherence}. 

In order to reduce the presence of TLSs in the junction, we deposit the Al/AlO$_{x}$ stack in a UHV chamber (Plassys) with a base pressure of 3$\times 10^{-10}$ Torr. Furthermore, the chamber is designed to avoid hydrocarbon contamination by utilizing high purity stainless steel lines for oxygen gas introduction and venting. The junctions fabricated in the UHV chamber generally yield qubits with higher $T_\textrm{2E}$, with an average $T_\textrm{2E}$ equal to $(1.2 \pm 0.4) T_{1}$ (Fig.~\ref{fig3:CPMG}(a,b)), and there is no significant difference in the span of fluctuations between HV and UHV devices, despite the higher average coherence times for qubits with UHV deposited junctions.

The improved junctions still exhibit device-to-device variation in $T_\textrm{2E}$. To investigate the nature of major sources of noise for qubits with junctions deposited in UHV, we select a qubit (qubit 2 in Table ~\ref{table:T1s_CL}) with $T_\text{2E}$ = 0.4 $T_1$ and use a Carr-Purcell-Meiboom-Gill (CPMG) dynamical decoupling pulse sequence to extract the corresponding noise spectral density \cite{CPMG_Cywinski2008, CPMG_Dwyer2022, PRAViola}. We note that this qubit is fabricated with a separate drive line to enable high Rabi frequency driving. We examine the fitted $T_{2,\text{CPMG}}$ decay for a varying number of CPMG pulses (Fig.~\ref{fig3:CPMG}(c)). For pulse sequences with $N < 100$ $\pi$-pulses, $T_{2,\text{CPMG}}$ remains constant at a value of 0.2~ms. For higher order pulse sequences, $T_{2,\text{CPMG}}$ increases as $N^{0.6}$ and approaches 2$T_{1}$ ($T_{1}$ = 0.56 ms). 

Using this CPMG data, we reconstruct the noise spectral density (Fig.~\ref{fig3:CPMG}(d)), following the method outlined in \cite{ Bylander2011Noise}. To isolate dephasing contributions, we remove the relaxation-induced decay ($T_1$ effects) from the data (see supplementary material). The noise spectral density is flat at low frequencies and then decreases beyond a cutoff frequency that is similar to the linewidth of the qubit's readout resonator,  $\kappa / 2\pi = 288$ kHz (dashed gray line). To explore the connection between this cutoff frequency and the resonator linewidth, we fit the noise spectrum to a model in which dephasing arises from coupling to thermal photons in the resonator~\cite{Yan2016flux}:
\begin{equation}
    S(\omega) = \frac{(2\chi)^22\eta \bar{n}\kappa}{\kappa^2 + \omega^2} + B
    \label{eqn: sw}
\end{equation}
where $\chi$ is the dispersive shift of the qubit,  $\eta = \kappa^2/(\kappa^2 + (2\chi)^2)$, and $\bar{n}$ is the mean thermal photon number. We group the numerator as a single constant $A$, and use $A$, $\kappa$, and $B$ as fit parameters. We note that this function differs from a Lorentzian function by a factor of two in $\kappa$, arising from the distinct photon distribution of thermal photons compared to coherent drives~\cite{thermalphotonDephasing}. The fit yields a value for $\kappa / 2\pi = 299 \pm 8$ kHz which is consistent with the measured linewidth of this resonator, suggesting that a significant noise source is photon shot noise arising from thermal photons in the readout resonator~\cite{Schuster2005ac, Yan2016flux}. For the Stark shift per photon in this device, $(2\chi)/2\pi = 300$ kHz, we use Equation~\ref{eqn: sw} to approximate the average thermal photon population in the resonator as $\bar{n} = 0.005$, corresponding to an effective temperature of 61 mK. 

The materials improvements that we describe here do not require any architectural changes to the 2D transmon qubit and are readily translated to existing control schemes. To demonstrate this, we perform single qubit randomized benchmarking (RB)~\cite{Knill2008RB} on qubit 2 in Table~\ref{table:T1s_CL}. Single qubit gate fidelities depend on the duration of the pulse; as the pulse length increases, qubit decoherence during the pulse leads to errors. As the pulse length decreases, the probability of leakage errors outside of the computational subspace increases ~\cite{WoodLeakage2018}. The high-coherence Ta-on-Si platform allows us to achieve high gate fidelities with limited optimal control. We use a Gaussian enveloped microwave pulse modified only by a first-order derivative-removal-by-adiabatic-gate (DRAG) scheme to mitigate leakage errors ~\cite{Gambetta2011DRAG, Chow2011DRAG} (see supplementary material). 

\begin{figure}
   \centering
  \includegraphics[width=70mm]{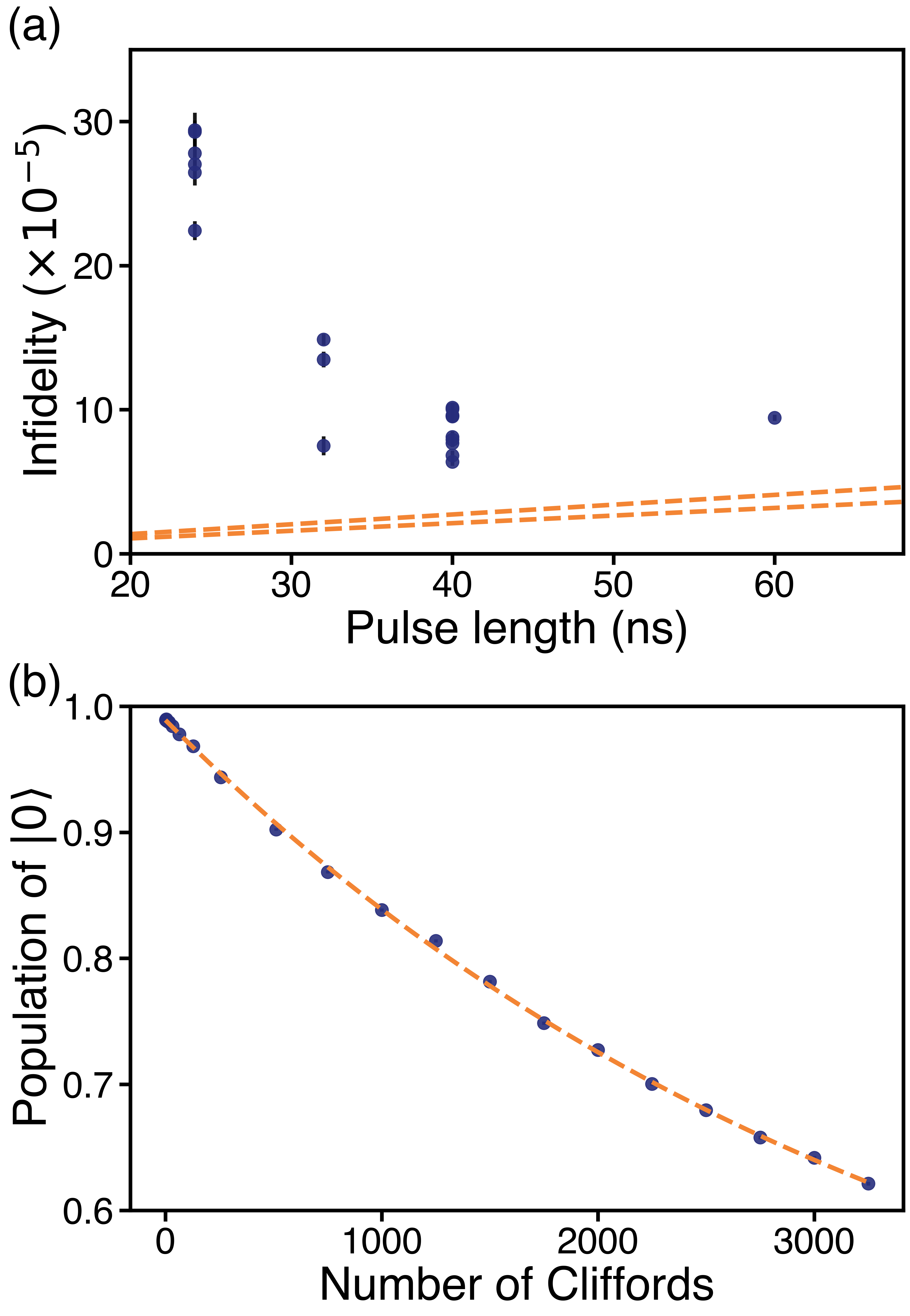}
  \caption{\textbf{High Single Qubit Gate Fidelity}: 
  (a) Average gate infidelity as a function of $X_{\frac{\pi}{2}}$ pulse length. The dashed lines represent the estimated fidelity limit imposed by the measured $T_1$, with temporal fluctuations of one standard deviation around the mean value. For long pulse lengths, the average gate infidelity is within $6 \times 10^{-5}$ of the estimated lifetime limit. At shorter pulse lengths, leakage errors dominate, which increases gate infidelity. The gate error is minimized for a 40~ns pulse. 
  (b) Return probability to the initial state as a function of the number of Clifford gates for a $X_{\frac{\pi}{2}}$ pulse length of 40~ns. For robust statistical results, we generate 100 different random sequences and measure each sequence 1000 times. The return probability decays exponentially with number of Clifford operations, and we calculate an average error per gate of $(6.4 \pm 0.3) \times 10^{-5}$.
  }

   \label{fig4:RB}
\end{figure}

We investigate the tradeoff between gate speed and coherence limits by performing RB at various $\pi/2$-pulse lengths ranging from 24 ns to 60 ns (Fig.~\ref{fig4:RB}(a)). Over the duration of the experiment (12 hours), $T_1$ has a mean value of 0.56~ms with a standard deviation of 0.07~ms. Therefore the fidelity limit imposed by $T_1$ fluctuates slightly in time (dashed yellow lines in Fig.~\ref{fig4:RB}(a)). We observe that for gate times of 60 and 40 ns the gate error exceeds this limit by an offset of $6\times 10^{-5}$ (see supplementary material). We attribute the excess error to temporal drifts in the control electronics and spectral diffusion of the qubit frequency. For shorter pulse lengths, the error per gate (EPG) increases with decreasing pulse length, suggesting that leakage errors are the dominant source of gate infidelity. The smallest observed average EPG is $(6.4 \pm 0.3) \times 10^{-5}$ with a 40 ns pulse length (Fig.~\ref{fig4:RB}(b)). Our highest average gate fidelity is in line with state-of-the-art gates on transmon qubits\cite{Li2023Error} despite using only straightforward control optimization, thus leaving room for further improvement~\cite{Eric2024reducing}. 


We have demonstrated 2D transmons fabricated from Ta-on-high-resistivity Si with lifetimes and coherence times exceeding 1~ms, and quality factors up to $2.5 \times 10^{7}$. The rationale for choosing and optimizing this material system was based on a systematic understanding of the dominant losses in transmon qubits: we utilize Ta as a metal because of its robustness to aggressive cleaning so that we can remove contamination while preserving the properties of its oxide to minimize surface losses, and we fabricate devices on high-resistivity Si to eliminate the bulk dielectric loss associated with sapphire. We then optimized the junction deposition conditions to avoid contamination that leads to decoherence. 

The present results open the door to several broad directions for near-term improvements in qubit performance, informed by measurements of the major contributions to loss and noise in this material platform. First, a major contribution to decoherence is excess photons in the readout resonator. Optimizing qubit parameters to reduce sensitivity to resonator photons or introducing new readout schemes that suppress dephasing~\cite{YoshimiReadoutPhotons, ZhangReadoutThermal} should lead to immediate improvements in the coherence times. Second, we have shown that high resistivity Si has low dielectric loss, but the losses at millikelvin temperatures are unrelated to free carriers. Instead, resistivity is likely a proxy for material purity or quality; ultrahigh-sensitivity materials analysis such as secondary ion mass spectrometry could potentially reveal the microscopic sources of loss, enabling further materials improvement. Third, dielectric loss in the current devices is now dominated by surface TLSs. Since this surface TLS-related loss tangent is the same for different Ta thin film crystal structures (from polycrystalline to single crystal)~\cite{crowley2023disentagling}, and is the same for Ta-on-sapphire and Ta-on-Si, we conclude that the dominant source of TLS loss is the amorphous tantalum oxide. Avoiding this oxide by noble metal encapsulation~\cite{chang2025eliminating} should remove this source of loss. Finally, the improvements in coherence time that we achieve through low-contamination Josephson junction deposition point to optimization of junction materials and fabrication as a major direction for future qubit improvement.

Beyond improving the average lifetimes and coherence times it will also be important to tackle sources of fluctuations in these parameters over time, as such fluctuations will limit the performance of large-scale processors. The temporal fluctuations of qubit $T_1$ and $T_\text{2E}$ are a large fraction of their average value, as quantified by mean $T_1$ and $T_{\text{2E}}$ spans of 36\% and 23\% respectively. This span is similar to that observed in previous generations of qubits, despite previous qubits having significantly lower overall coherence \cite{McRae2021reproducible,Klimov2018fluctuations}. Further study of the microscopic sources of these fluctuations and their interactions with the environment will be a critical area of future study.

More broadly, the materials improvements we outline here do not require any new qubit architectures and are immediately translatable to large-scale processors. Si in particular is an attractive material for scaling up fabrication, as it allows us to leverage the major global investment in Si growth, fabrication, metrology, and back-end-of-line processing that has been developed for the semiconductor industry~\cite{van2024advanced}. Furthermore, the millisecond-scale coherence of Ta-on-Si qubits will enable their use as an important metrology tool in understanding subtle sources of loss in large-scale processors. Their long lifetimes and coherence times make them a sensitive probe of cross-talk~\cite{Vinay2022Crosstalk}, nonequilibrium quasipartcles~\cite{Connolly_Quasiparticle}, vortex-motion induced losses~\cite{bahrami_vortex}, and correlated errors~\cite{CosmicRayGoogle, CosmicRays_Harrington}.

\bibliography{bib}

\textbf{Data availability:}
The data that support the findings of this study are available from the corresponding author upon reasonable request.

\textbf{Code availability:}
The code related to the data analysis of this study is available from the corresponding author upon reasonable request.

\begin{acknowledgments}
We gratefully acknowledge helpful conversations with Michel Devoret, Yu Chen, Alex Opremcak, George Sterling, Matt Reagor, Lev Ioffe, Lara Faoro, Jeff Thompson, Jared Rovny, Lev Krayzman, Parth Jatakia, Jake Bryon, and Kevin D. Crowley. We also acknowledge Esha Umbarkar and Maxwell Lin for initial work on Ta deposition on silicon. This work was primarily supported by the U.S. Department of Energy, Office of Science, National Quantum Information Science Research Centers, Co-design Center for Quantum Advantage (C$^{2}$QA) under Contract No. DESC0012704. This research was partially supported by Google Quantum AI. The authors acknowledge the use of Princeton’s Imaging and Analysis Center (IAC), which is partially supported by the Princeton Center for Complex Materials (PCCM), a National Science Foundation (NSF) Materials Research Science and Engineering Center (MRSEC; DMR-2011750), as well as the Princeton Micro/Nano Fabrication Laboratory. We also acknowledge MIT Lincoln Labs for supplying a traveling wave parametric amplifier.

Princeton University Professor Andrew Houck is also a consultant for Quantum Circuits Incorporated (QCI). Due to his income from QCI, Princeton University has a management plan in place to mitigate a potential conflict of interest that could affect the design, conduct and reporting of this research.
\end{acknowledgments}

\newpage

\setcounter{figure}{0}
\renewcommand{\thetable}{\textbf{\arabic{table}}}%

\renewcommand{\figurename}{\textbf{Extended Data Fig.}}

\renewcommand{\thefigure}{\textbf{\arabic{figure}}}%
\clearpage
\section*{Extended Data}

\section{Methods} 
\label{Methods}
$\alpha$-Ta films are deposited on 4 inch high-resistivity silicon wafers from Siegert Wafer GmbH and Topsil. The Siegert intrinsic silicon wafers are grown using the float zone growth method, are double-side polished, and have a resistivity of $>$20~k$\Omega$-cm. The Topsil high-purity silicon wafers (undoped) are also grown using the float zone growth method, are single-side polished, and have a resistivity between 20 and 99~k$\Omega$-cm. Wafers from both vendors have an out-of-plane $<$100$>$ $\pm$ 0.5$^{\circ}$ orientation and a thickness of 525 $\pm$ 25~$\mu$m. Each wafer is cleaned in a piranha solution (2:1  H$_{2}$SO$_{4}$:H$_{2}$O$_{2}$) for 20 minutes, followed by three deionized (DI) water. To remove the surface oxide, the wafer is then placed in a beaker of 10:1 DI water:49\% HF  for two minutes, followed by three consecutive rinses in DI water. The wafer is finally rinsed in a beaker of 2-propanol and blowdried with N$_{2}$ gas.

The Si wafer is transferred to the load lock of an ultra-high vacuum (UHV) DC-magnetron sputtering system (AJA Orion 8) within 15 minutes of the BOE step to avoid oxide regrowth. As described in the main text, we use a Ta deposition with the shutter closed to reduce the base pressure after transferring the sample from the loadlock. For the getter step, Ta is sputtered at room temperature with 300 W of DC sputtering power and 3 mTorr of argon pressure for 500 - 1000 seconds. The Si wafer is then heated to 600 - 650\,$^{\circ}$C for 20 minutes to degas the chamber and holder and to remove any residual hydrocarbons from the surface of the wafer. The Ta deposition occurs at 600 - 650\,$^{\circ}$C at a rate of  0.3\,nm/s, with 300 W of DC sputtering power and 3\,mTorr of argon pressure. The wafer is cooled to room temperature over several hours to ensure that the wafer temperature reaches 25~$^{\circ}$C before unloading. 

The Ta-on-Si wafer is then diced into smaller pieces for the fabrication process. Photoresist (AZ1518) is spun on the surface of the wafer to protect the film prior to dicing into either $25\,\text{mm}$ or $10\,\text{mm}$ squares. The photoresist is then stripped using Remover PG for an hour at 80~$^{\circ}$C on a hot plate, then sonicated in consecutive beakers of acetone and 2-propanol, and then dried with N$_{2}$ gas.

The Ta-on-Si pieces undergo a Hexamethydisilazine (HMDS) vapor step at 148\,$^{\circ}$C to promote surface adhesion. AZ1518 photoresist is then spun on the wafer at 4000\,rpm for 40\,s with a 1000\,rpm/s ramp rate, followed by a soft bake at 95\,$^{\circ}$C for a minute, exposed (Heidelberg DL66+ laser writer), followed by a two minute post-exposure bake at 110\,$^{\circ}$C, and then a photoresist development step with AZ300MIF for 90 seconds and a 30\,s rinse in DI water. 

Following development, the Ta-on-Si films are etched via a chlorine reactive ion etching (RIE) process using a PlasmaTherm Takachi SLR inductively coupled plasma (ICP) etcher (5.4\,mTorr, 500~W ICP, 50~W RF bias, 5~sccm each of Cl$_2$ and Ar, etch rate = 250 nm / min).

After the etch step, the photoresist (AZ1518) is stripped using Remover PG for an hour at 80\,$^{\circ}$C on a hot plate, sonicated in consecutive beakers of acetone and 2-propanol, and then dried with N$_{2}$ gas. The devices are further cleaned in 30 mL of piranha solution  (2:1  H$_{2}$SO$_{4}$:H$_{2}$O$_{2}$) for 20 minutes, followed by three consecutive rinses in beakers of DI water. We reduce the thickness of the Ta native oxide layer via a buffered oxide etch process (10:1 BOE) for 20 minutes, followed by three consecutive rinses in DI water followed by 2-propanol, and dried with N$_{2}$ gas.

For junction fabrication, the chip then undergoes HMDS vapor treatment before spinning a bi-layer electron beam resist stack. First, MMA (MMA (8.5) MAA EL 13) is spun at 5000 RPM for 70 seconds and hard baked for 2 minutes at 175 $^{\circ}$C.  Then, PMMA (AR-P 672.045 PMMA 950K) is spun at 4000 RPM for 68 seconds and hard baked for 5 minutes at 175 $^{\circ}$C). We then spin an  anti-charging layer (Discharge x4 polymer) at 4000rpm for 40\,seconds. The resist is patterned using electron-beam lithography (Raith EBPG 5150+). Before developing, we remove the anti-charge layer with DI-water for 10 seconds.  The resist development is done in a MIBK/IPA 1:3 developer at room temperature for 50 seconds followed by 10 seconds in IPA, and dried with N$_{2}$ gas. The device is then mounted in an electron-beam evaporation system (Plassys MEB550S and Plassys UHV) for double-angle deposition with an undercut bilayer resist. Argon ion milling is used to remove residual and tantalum oxide before aluminum deposition to enable good electrical contact. A Ti getter step is performed before junction deposition to lower the chamber pressure to 3$\times 10^{-10}$ mbar. The junction evaporation step is performed with the following parameters: the first layer is deposited with a 40$^{\circ}$ angle and thickness of 20~nm in the UHV chamber, followed by a 15 minute oxidation step at 50 mbar of pure O$_{2}$ in a separate oxidation chamber. Then, the second layer is deposited at a similar deposition angle with 70~nm thickness in a UHV chamber, followed by a 20 minute oxidation step at 10 mbar of pure O$_{2}$.

After evaporation, liftoff is performed using Remover PG for 3 hours at 120 $^{\circ}$C on a hot plate, then sonicated in consecutive beakers of acetone and 2-propanol, and then dried with N$_{2}$ gas.

After fabrication, the device is wire bonded to create electrical contacts from the chip to the PCB. Additionally, we use on-chip wirebonds to connect ground planes and avoid packaging modes. The device is enclosed in a commercial sample holder: Quantum Machines QCage.24 Chip Carrier.

\begin{figure}
    \centering
    \includegraphics[width=0.7\linewidth]{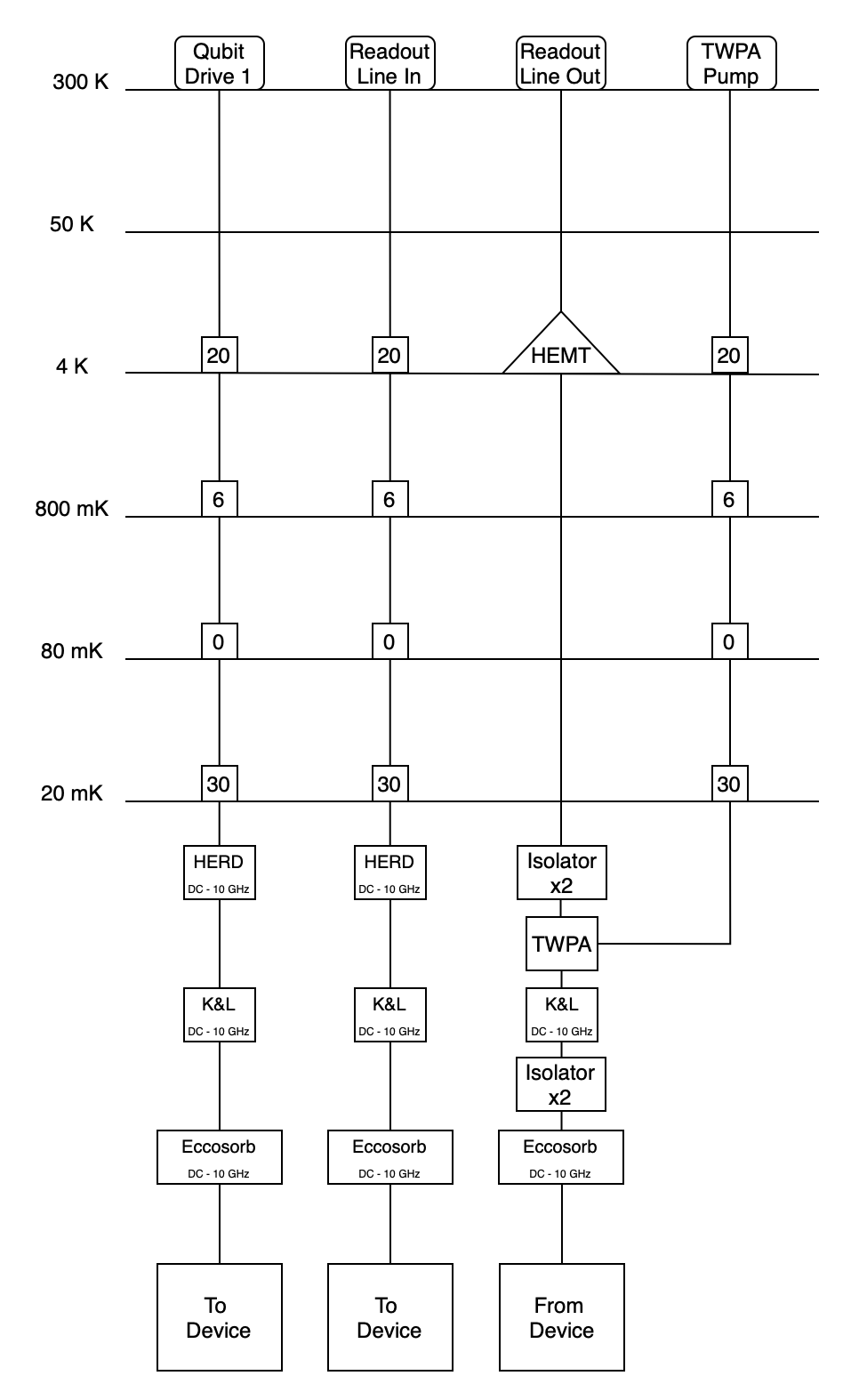}
    \caption{Wiring diagram for dilution refrigerator.  Numbers in boxes indicate attenuation in dB.}
    \label{fig:dil-fridge}
\end{figure}

\begin{figure}
    \centering
    \includegraphics[width=0.5\linewidth]{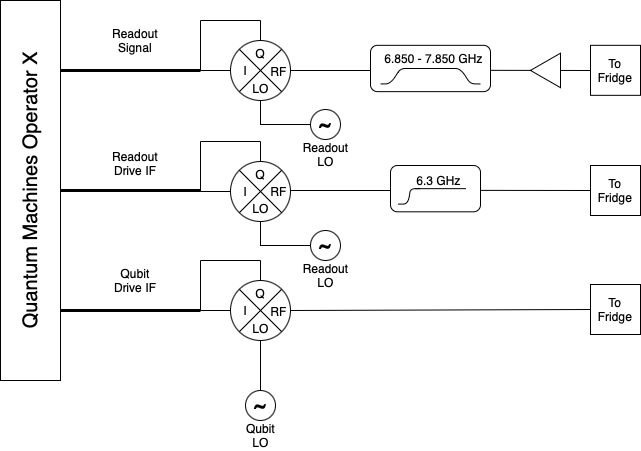}
    \caption{Schematic of room temperature electronics used when measuring with the Quantum Machines Operator X.}
    \label{fig:OPX}
\end{figure}

The device is mounted in a Bluefors XLD400 or LD400 dilution refrigerator with a base temperature of 13 mK. A wiring diagram with the corresponding filtering and attenuation can be seen in Extended Data Fig.~\ref{fig:dil-fridge}. For room-temperature electronics, time domain measurements were conducted with both a QICK-controlled Xilinx RFSoC ZCU216~\cite{stefanazzi_qick_2022} and a Quantum Machines Operator X (OPX). The RFSoC board directly synthesized the qubit (2-4 GHz) and resonator (6.5-7.5 GHz) drive and readout tones, reducing the need for external mixers. Extended Data Fig.~\ref{fig:OPX} shows the wiring and up conversion for the OPX. Data for $T_1$ and $T_{2\text{E}}$ measurements was primarily taken with the RFSoC while measurements for single-qubit RB and CPMG were conducted with the OPX. 

\begin{figure}
   \centering
  \includegraphics[width=0.7\textwidth]{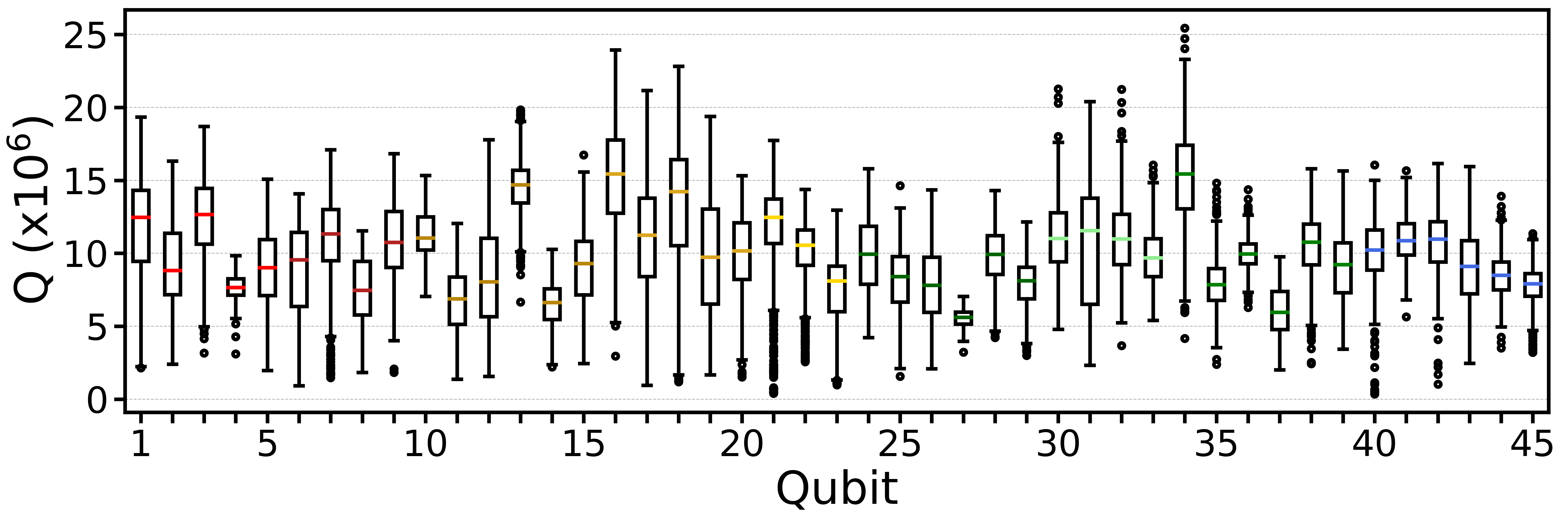}
  \caption{Comparison of $Q$ for 45 transmon qubits across 9 chips. The box marks the 25$^\text{th}$ and 75$^\text{th}$ quartile and the colored line marks the median of the dataset. Qubits fabricated on the same chip are grouped by color.}
   \label{SI:QAvg}
\end{figure}

\begin{figure}
   \centering
  \includegraphics[width=0.5\textwidth]{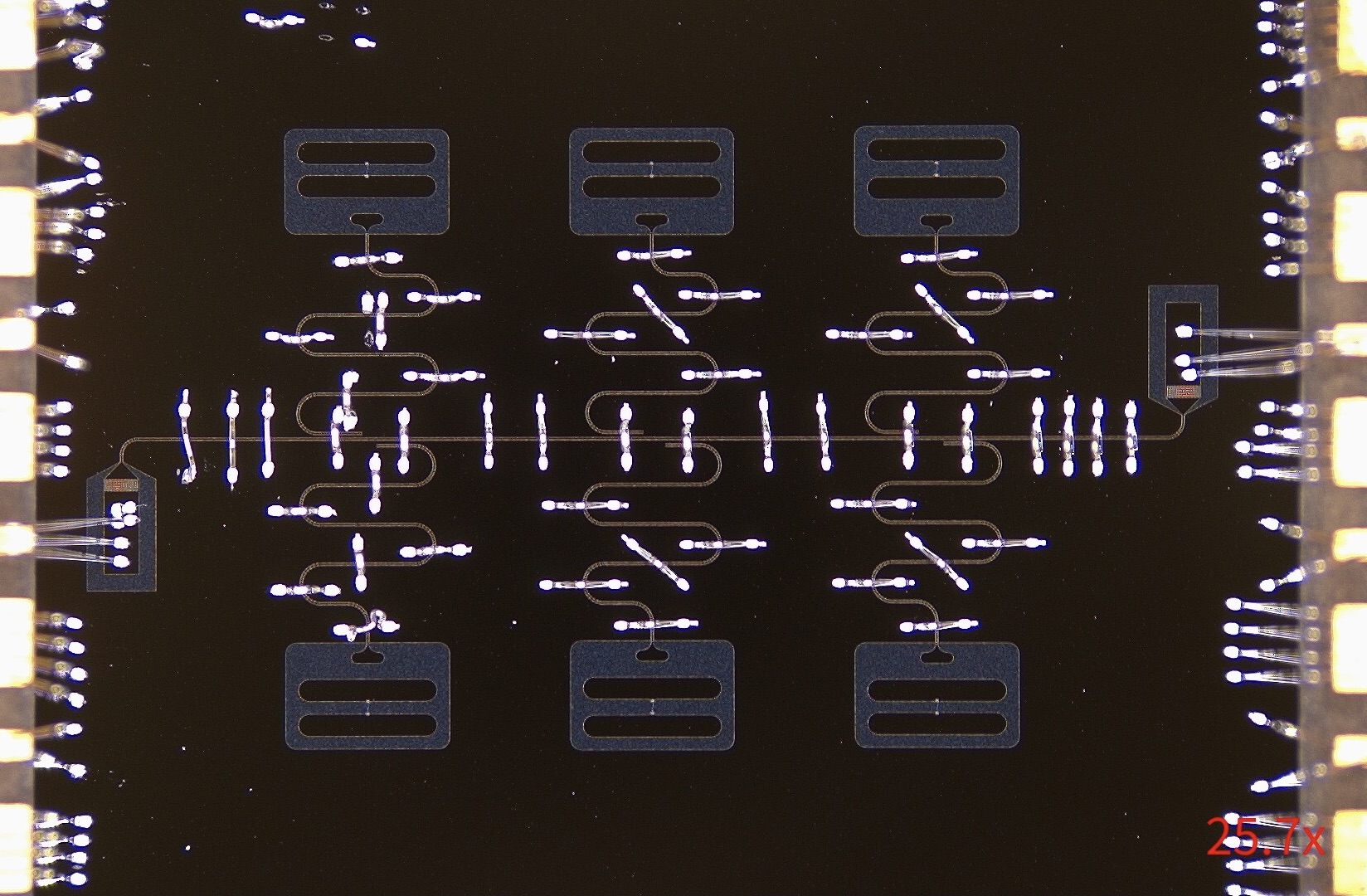}
  \caption{\textbf{Transmon device pattern}.
 Optical image of the qubit chip wire-bonded in a QCage.24 Chip Carrier. The 7~mm $\times$ 7~mm pattern comprises six transmons, each capacitively coupled to an independent readout resonator which is inductively coupled to a central feedline that also acts as a Purcell filter. The measured device parameters for these qubits are shown in Table~\ref{table:devparams}, with a simulated $T_{1,p}$ that is the $T_{1}$ due to Purcell decay. The qubit parameters were designed in AWR Microwave Office.}

   \label{fig:devicepattern}
\end{figure}

The measurements were conducted across 9 different chips, each containing 6 qubits in the configuration shown in Extended Fig.~\ref{fig:devicepattern}. Of those, we measure and report repeated relaxation decay constants, $T_1$, of 45 viable qubits. Each one was measured for at least three days, providing a time-averaged $T_1^\text{avg}$, maximum $T_1^\text{avg}$ over time, and their corresponding quality factors $Q = \omega \times T_1$, where $\omega / 2 \pi$ is the qubit frequency. The $T_1$ and $Q$ data for our qubits are shown in Table~\ref{table:T1s} and Extended Fig.~\ref{SI:QAvg}. The remaining qubits either had junctions that were shorted or were in parameter regimes with very poor driveability. An additional set of 12 qubits across two chips were measured (Table~\ref{table:T1s_CL}). These qubits differ from those in Table~\ref{table:T1s}; they included separate drive lines that were used for the CPMG and RB experiments. Typical qubit and readout parameters are shown in Table~\ref{table:devparams}. 


\section{Extracting $Q_{\text{TLS}}$ from resonator measurements}

All resonators are coupled to a common feedline in a hanger configuration. The complex transmission coefficient ($S_{21}$) is measured around resonance with a vector network analyzer (VNA) and is modeled as

\begin{equation}
\label{eq:S21}
S_{21}(f) = A e^{i(\theta + 2\pi f \tau)} \left( 1 - \frac{(Q_l / |Q_c|) e^{i \phi}}{1 + 2 i Q_l \left( \frac{f - f_0}{f_0} \right)} \right)
\end{equation}
where $f$, $f_0$, $Q_{l}$, $|Q_{c}|$, $\phi$, $A$, $\alpha$, and $\tau$ are the probe frequency, resonant frequency, loaded quality factor, the coupling quality factor, a phase factor that accounts for impedance mismatch between the resonator and the feedline, the background amplitude, phase due to overall gain or attenuation in the measurement chain, and the cable delay.

\begin{figure}
   \centering
  \includegraphics[width=140mm]{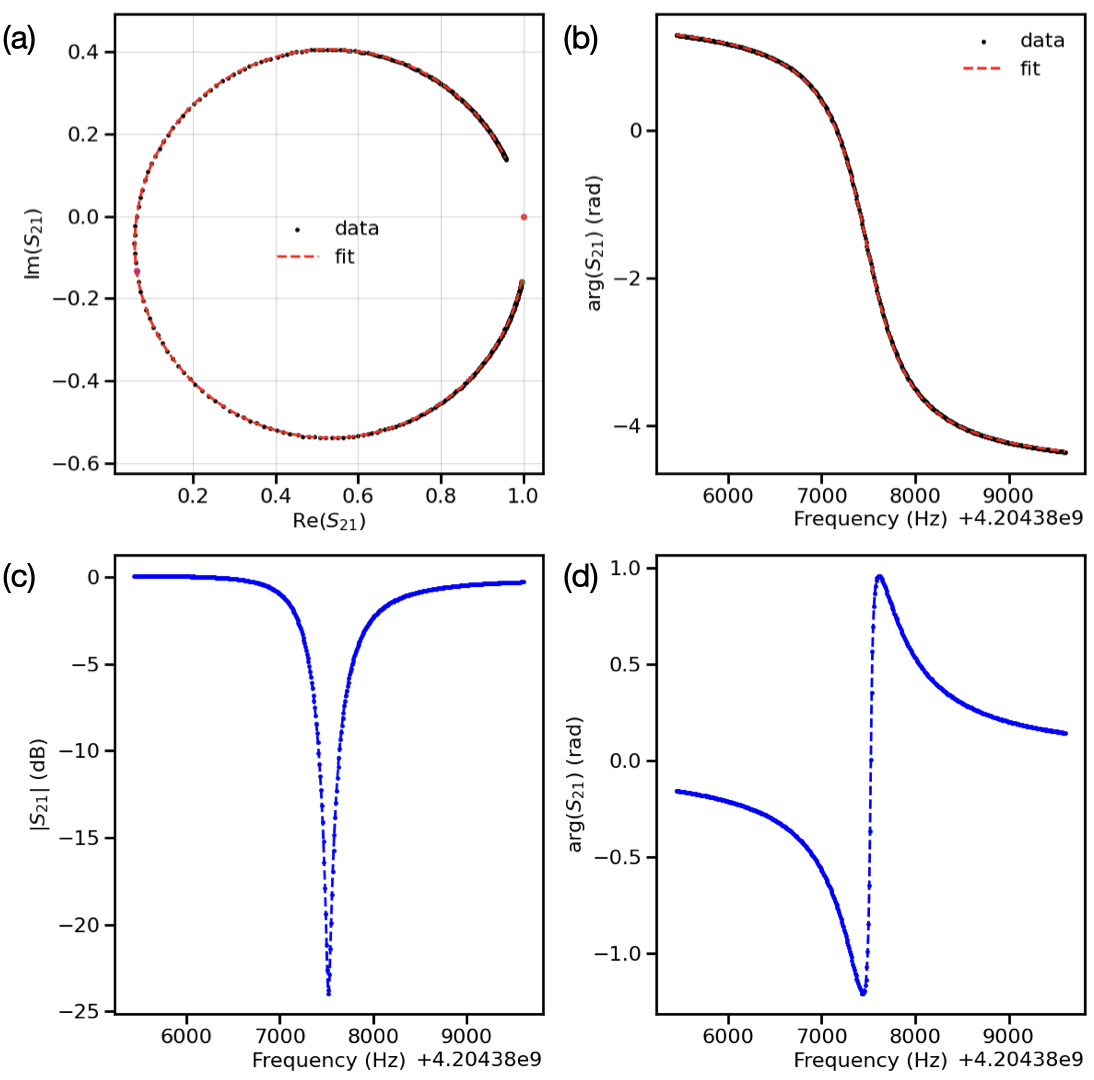}
  \caption{\textbf{Resonator spectroscopy}. (a) Resonance circle obtained after accounting for the background amplitude, phase, and cable delay, which normalizes the off-resonant response to unity. This circle is fitted to obtain its center $c$ and radius $r_0$. (b) Frequency dependence of the principal argument of the resonance circle centered at the origin. This curve is fitted to obtain $Q_l$ and $\omega_0$. $|Q_c|$ and $\phi$ are calculated from the geometric relations between $Q_l$, $c$, and $r_0$. The amplitude and phase projections of the resonance circle are depicted in (c) and (d) respectively.}
   \label{fig:rescircle}
\end{figure}

\begin{figure}
   \centering
  \includegraphics[width=0.7\textwidth]{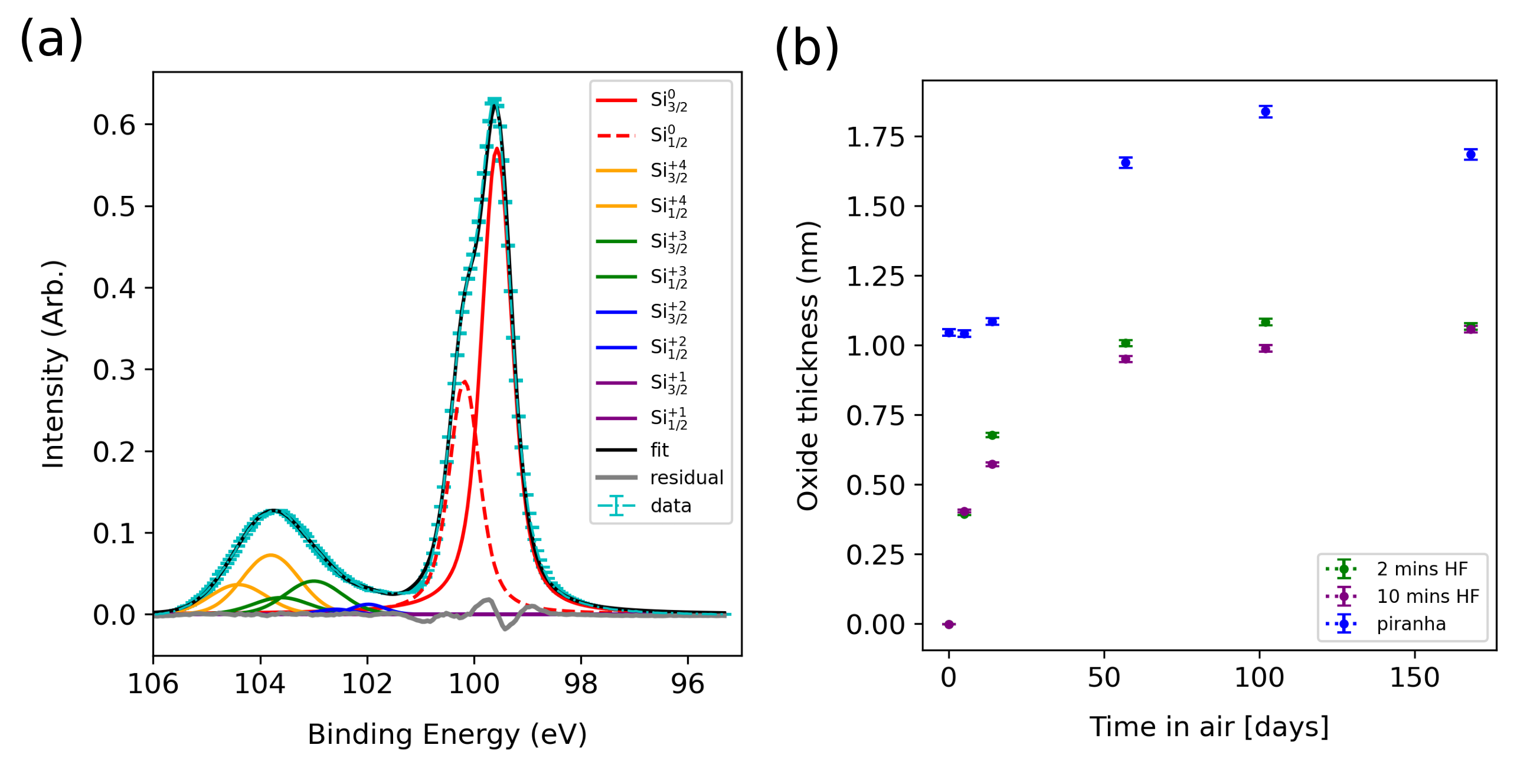}
  \caption{\textbf{SiO$_{x}$ Regrowth Rate}: (a) Si2p XPS spectrum showing a pair of spin-orbit split peaks corresponding to bulk Si (Si$^0$) and a broad feature associated with the oxide. We fit the spectrum to several pairs of oxide peaks corresponding to SiO$_2$ (Si$^{4+}$) and suboxides (Si$^{3+}$, Si$^{2+}$, Si$^{1+}$) to calculate a total oxide thickness of  $1.84 \pm 0.02$ nm after exposure to ambient conditions for 102 days. (b) The total oxide thickness calculated from the Si2p XPS spectrum for Si wafers left in ambient conditions after treatment in HF for 2 minutes (green) and 10 minutes (purple), as well as piranha (blue). The oxide growth rate varies by surface preparation method.}

   \label{SI:SIOx}
\end{figure}

\begin{figure}
   \centering
  \includegraphics[width=0.7\textwidth]{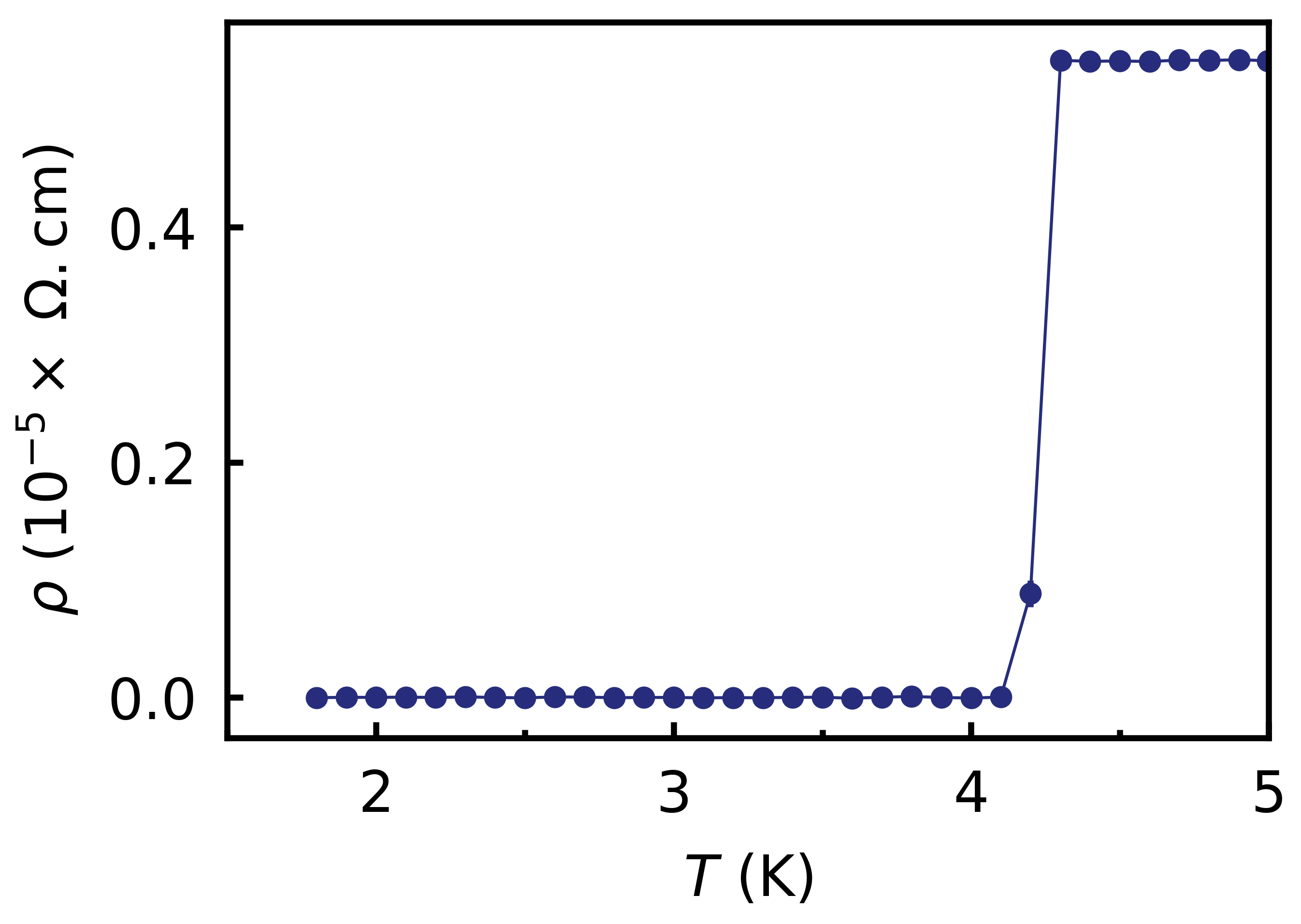}
  \caption{Resistivity as a function of temperature for Ta-on-Si films at zero applied magnetic field. The transition temperature is 4.2 $\pm$ 0.1~K, consistent with $\alpha$-phase Ta.}

   \label{Ta-characterization}
\end{figure}

The internal quality factor is calculated from $Q_{\mathrm{int}}^{-1} = Q_l^{-1} - |Q_c|^{-1}\cos \phi$ \cite{khalil2012analysis}. Equation \ref{eq:S21} traces a circular arc in the complex plane (Figure \ref{fig:rescircle}) and is fitted using the procedure described in \cite{probst2015efficient}.

We then measure $Q_\text{int}$ as a function of temperature and microwave power to estimate $Q_{\text{TLS,0}}$. Figure~\ref{SI:waterfall} shows the resonator's internal quality factor, $Q_\text{int}$, as a function of temperature for a wide range of microwave powers. Similar to our previous work~\cite{crowley2023disentagling}, the measured $Q_\text{int}$ for Ta-on-Si resonators can be adequately modeled with our three-component loss model: TLSs ($Q_{\text{TLS}}$), equilibrium quasi-particles ($Q_{\text{QP}})$, and a separate power and temperature-independent loss ($Q_{\text{other}}$) term,
\begin{equation}
    \frac{1}{Q_{\text{int}}} = \frac{1}{Q_{\text{TLS}}(\bar{n}, T)} + \frac{1}{Q_{\text{QP}}(T)} + \frac{1}{Q_{\text{other}}}
\end{equation}
where $Q_{\text{TLS}}(\bar{n}, T)$ and $Q_{\text{QP}}(T)$ are:
\begin{equation}
    Q_{\text{TLS}}(\bar{n}, T) = Q_{\text{TLS,0}} \frac{\sqrt{1+(\frac{\bar{n}^{\beta_{2}}}{DT^{\beta_{1}}})\tanh(\frac{\hbar\omega}{2k_{B}T})}}{\tanh(\frac{\hbar\omega}{2k_{B}T})}
\end{equation}
and
\begin{equation}
    Q_{\text{QP}}(T) = A_{\text{QP}}\frac{e^{\Delta_0/k_{B}T}}{\sinh(\frac{\hbar\omega}{2k_{B}T})K_{0}(\frac{\hbar\omega}{2k_{B}T})}
\end{equation}
where $\omega$, $T$, $\bar{n}$, $Q_{\text{TLS,0}}$are the center angular frequency of the resonator, temperature, the intracavity photon number, and inverse linear absorption from TLSs, respectively. $D$, $\beta_{1}$, and $\beta_{2}$ are fit parameters characterizing TLS saturation. The variables $Q_{\text{QP}}$, $A_{\text{QP}}$, $\Delta_{0}$, $K_{0}$, $k_{B}$, and $\hbar$ are the quality factor due to equilibrium quasi-particle loss, an overall amplitude proportional to the kinetic inductance ratio, the superconducting gap, the zeroth-order modified Bessel function of the second kind, the Boltzmann constant, and the reduced Planck constant, respectively. At millikelvin temperatures and single-photon powers, the relevant conditions for a 2D transmon, the dominant loss source is TLSs. Therefore, we extract the parameter that estimates the linear absorption due to TLSs, $Q_{\text{TLS,0}}$.


\section{X-ray Photoelectron Spectroscopy (XPS)}

All XPS data were taken using a Thermofisher K-Alpha or a Nexsa G2 X-ray Photoelectron Spectrometer System, both using monochromated Al K$\alpha$ = 1486 eV X-ray sources. XPS spectra of the Si2p core transition were taken in 0.05 eV steps and averaged over 10 scans with a spot size of 400 $\mu$m and a dwell time of 50 ms.

After data collection, each Si2p spectrum is shifted in binding energy to center the Si$^{0}$ peak at 99.4 eV~\cite{YUBERO2000spec} and normalized by the total area (intensity) under each spectrum. Then each spectrum is fitted with a Shirley background from 95 - 110 eV and fit to a multicomponent model containing Lorentzian lineshapes for metallic peaks (Si$^{0}$) and Gaussian lineshapes for other oxidation states (Si$^{1+}$, Si$^{2+}$, Si$^{3+}$, Si$^{4+}$). Each Si2p peak consists of two spin-orbit peaks Si2p$_{1/2}$ and Si2p$_{3/2}$ with a 2:1 intensity ratio and spin-orbit splitting of 0.6 eV~\cite{YUBERO2000spec}. 

After fitting, the area of each component in the spectrum is calculated from the fitted parameters.  An area ratio of each component is then used to find the approximate thickness of the total oxide including all non-zero oxidation states. The area ratios of the metal and the oxide are then used to calculate the total oxide thickness using Eqn.~\ref{eqn:Strohmeier}~\cite{strohmeier1990ESCA}.

\begin{equation}
 d = \lambda_{ox} \sin \theta \ln(\frac{N_{m}}{N_{ox}} \cdot \frac{I_{m}}{I_{ox}} \cdot \frac{\lambda_{ox}}{\lambda_{m}} +1 )
\label{eqn:Strohmeier}
\end{equation}

In Eqn.~\ref{eqn:Strohmeier}, $\ I_{m}$ ($\ I_{ox}$) is the intensity (peak area) of the metal (oxide), $\ N_{m}$ ($\ N_{ox}$) the volume density of metal atoms in the metal (oxide), and $\lambda_{m}$ ($\lambda_{ox}$) the inelastic mean free path of photoelectrons in the metal (oxide). $\theta = 90^{\circ}$ is the angle of photoelectron ejection with respect to the surface. Values for $\ I_{m}$ and $\ I_{ox}$ were taken from~\cite{Powell2001measurement} and the ratio of $\ N_{m}$ to $\ N_{ox}$ was calculated to be 2.139.


\section{Noise Spectroscopy}

We investigate the noise mechanisms of our qubits using CPMG decoupling schemes. These consist of two $X_{\pi/2}$ pulses with a series  $(\tau/2 -Y_{\pi}-\tau/2)^N$ between them. $N$ is the number of CPMG pulses applied and $\tau = t / N$ where $t$ is the total free evolution time between the $X_{\pi/2}$ pulses.

Following the methodology of \cite{Bylander2011Noise}, the qubit population under CPMG pulses is given as 
\begin{equation}
    P(t) = 0.5 + 0.5e^{-t/(2T_1)}e^{-\Gamma_pN}e^{-\chi_N(t)} \label{eqn:population}
\end{equation}
where $\Gamma_p$ is the coherence decay experienced while $Y_{\pi}$ pulses are applied. We extract this using an exponential fit on the qubit population $P(t = 0)$ for different numbers of decoupling pulses. We find $\Gamma_p$ = 1/13000 for our $\pi$-pulse lengths of 80 ns. $\chi_N(t)$ is the coherence integral given as:
\begin{equation}
\chi_N(t) = t^2 \int_0^{\infty}d\omega S(\omega) g_N(\omega, t) \label{eqn: chi_eq}
\end{equation}
\noindent and $g_N(\omega, t)$ is a frequency filter function that depends on the number of CPMG pulses given as:
\begin{equation}
    g_N(\omega, t) = \frac{1}{(\omega t)^2} | 1 + (-1)^{1 + N} e^{i\omega t} + 2\sum_{j = 1}^N (-1)^j e^{i\omega \delta_j t}\cos(\omega t_{\pi}/2)|^2
\end{equation}
\noindent with $\delta_j$ being the normalized position of the center of the $j$th pulse in between the two $X_{\pi/2}$ pulses, $t$ is the total evolution time, and $t_{\pi}$ is the length of each $X_{\pi}$ pulse. The filter function is centered at $\omega / 2\pi = N/2t$ and its width is given by the inverse of the total evolution time. Approximating it as a delta function with a given area, Equation~\ref{eqn: chi_eq} can be simplified to:
\begin{equation}
    \chi_N(t) = t^2S(\omega'(t))g_N(\omega'(t), t)\Delta w \approx  t S(\omega') g_N(\omega'(t), t)
\end{equation}

\
To extract $S(\omega)$, we first remove the effects from $T_1$ decay and dephasing during the decoupling pulses:
\begin{equation}
    P'(t) = 2\ (P(t) - 0.5)\ e^{t/(2T_1)}\ e^{\Gamma_pN}
\end{equation}
\noindent Then we extract $(\omega)$:
\begin{equation}
    S(\omega(N, t)) = -\ln(P'(t)) / (t * g_N(\omega(N, t)) 
\end{equation}
\noindent where $g_N(\omega(N, t))$ is numerically calculated.

\section{Gate Calibration}

\begin{figure}
    \centering
    \includegraphics[width=1\linewidth]{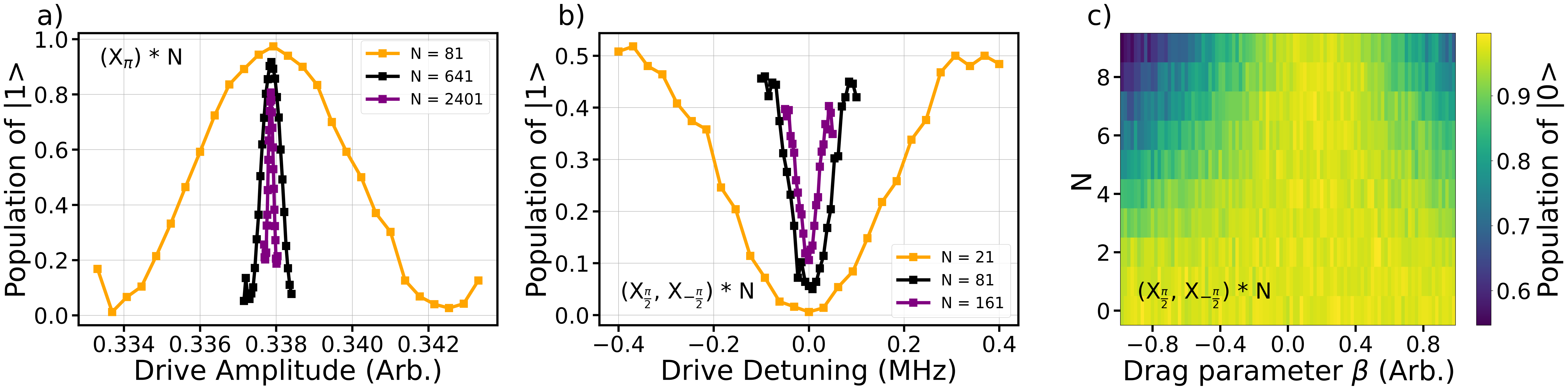}
    \caption{\textbf{Gate Calibration for Single Qubit Randomized Benchmarking} \textbf{a)} Excited state population versus drive amplitude. The inset sequence is used to calibrate the pulse amplitude of the $X_\frac{\pi}{2}$ in a Rabi-type experiment. Increasing  $N$ results in finer amplitude resolution. \textbf{b)} Excited state population versus drive detuning. The inset sequence is used to calibrate the drive frequency of the $X_\frac{\pi}{2}$. Increasing  $N$ results in finer frequency resolution. \textbf{c)} Population versus $N$ and DRAG parameter $\beta$. The inset sequence is used to calibrate DRAG for the $X_\frac{\pi}{2}$ pulse. At the optimal $\beta$, qubit population in $\ket{0}$ is highest for all $N$.} 
    \label{fig:GateCalib}
\end{figure}

\begin{figure}
   \centering
  \includegraphics[width=0.7\textwidth]{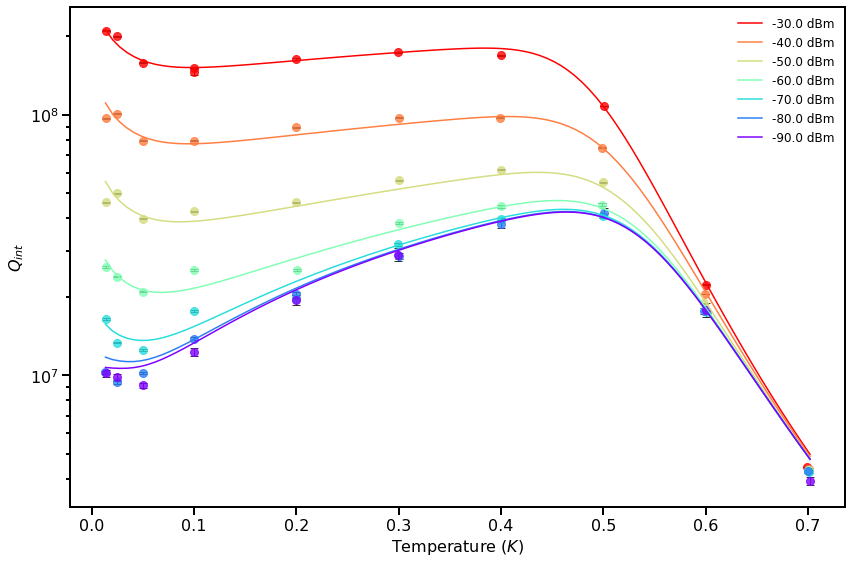}
  \caption{\text{Power and temperature measurements of resonator losses} $Q_\text{int}$ as a function of temperature and applied microwave power for a characteristic Ta-on-Si resonator. The traces are well separated at low temperatures and then collapse together and fall exponentially at high temperatures. The characteristic shape of the curves is fit to a model incorporating TLS loss and equilibrium quasiparticles~\cite{crowley2023disentagling}. Solid lines show the best fit to the dataset.}

   \label{SI:waterfall}
\end{figure}

We implement our $X_{\frac{\pi}{2}}$ rotation with a Gaussian waveform 
\begin{equation}
    \Omega(t) = \Omega_0(e^{-\frac{(t - t_g / 2)^2}{2\sigma^2}} - e^{-\frac{t_g^2}{8\sigma^2}})
\end{equation}
with $t_g$ being the gate length, $\sigma = t_g/4$ and $\Omega_0$ being the drive amplitude. We implement derivative removal by adiabatic gate (DRAG) to suppress leakage and phase errors with the envelope 
\begin{equation}
    \Omega_\text{DRAG}(t) = \Omega(t) - i \beta \frac{\dot{\Omega}(t)}{\alpha}
\end{equation} 
where $\alpha$ is the qubit anharmonicity and $\beta$ is the drag parameter that we sweep over to improve gate fidelity~\cite{Gambetta2011DRAG}\cite{Chow2011DRAG}. We calibrate our single qubit pulses with a series of Rabi-type line cut experiments. For drive amplitude calibration, we apply $N$ repeated $X_\pi$ pulses each composed of two $X_\frac{\pi}{2}$ pulses. For odd values of $N$, the proper amplitude will produce the highest population in state $\ket{1}$. Increasing $N$ amplifies rotation errors due to imperfect drives, allowing fine tuning of the drive amplitude. To calibrate the drive frequency, we perform a pseudo-identity operation consisting of $(X_\frac{\pi}{2}X_{-\frac{\pi}{2}})^N$. Increasing $N$ enhances errors caused by detuned drives, allowing us to calibrate frequency with increased precision. Examples of this calibration protocol are shown in Extended Data Fig. \ref{fig:GateCalib}. This protocol is similar to that used in Li et. al~\cite{Li2023Error}.

\section{Lifetime Limit of Gate Fidelity}
RB experiments consist of sequences of random Clifford operations followed by an inversion operation that returns the system to its initial state. As such, the RB sequence has the effect of a dynamical decoupling sequence. The Rabi frequency of our applied pulses exceeds 16~MHz, which is far beyond the noise cutoff frequency we observe in Fig.~\ref{fig3:CPMG}(d). Under the assumption that there is no latency between pulses, the RB sequence therefore decouples the pure dephasing $T_\phi$ of the qubit. This results in an effective coherence time that is $T_1$-limited ($T_{2, \text{RB}}$ = $2T_1$). We can model a purely $T_1$-limited system by the amplitude damping channel $\mathcal{E}_{AD}(\rho)$ acting on some state $\rho$ \cite{MikeAndIke}:
\begin{equation}
\begin{aligned}
    &\mathcal{E}_{AD}(\rho) =  A_0 \rho A_0^\dagger + A_1 \rho A_1^\dagger \\[10pt]
    &A_0 =
    \begin{bmatrix} 
        1 & 0 \\ 
        0 & \sqrt{1 - \gamma}
    \end{bmatrix}, \quad
    A_1 =
    \begin{bmatrix} 
        0 & \sqrt{\gamma} \\ 
        0 & 0
    \end{bmatrix}
\end{aligned}
\end{equation}
with the time-varying substitution for qubit $T_1$,
\begin{equation}
   \gamma = 1- e^{-\frac{t_g}{T_1}}.
\end{equation}
where $t_g$ is the gate pulse length. The average gate fidelity, $\bar{F}_{\mathcal{E}}$, for a channel with Kraus operators\{$K_i$\}, can be expressed as \cite{Wood2015TensorNetworks}:
\begin{equation}
    \bar{F}_\mathcal{E} = \frac{d +\sum_i| \text{Tr}[K_i]|^2}{d(d+1)}
\end{equation} 
We use $\bar{F}_{\mathcal{E}_{AD}}$ to calculate the coherence limit of our gate fidelity.

\begin{table*}
\centering
\caption{\label{table:T1s} Device parameters, lifetime, and coherence times for 45 Ta-on-Si qubits fabricated across 9 chips.}
\begin{ruledtabular}
\begin{tabular}{cccccccccc}
        Qubit     & Pitch  &   Deposition   &   Frequency   &   $T_{1}^\text{avg}$   &   $T_{1}^\text{max}$   &   $Q^\text{avg}$   &   $Q^\text{max}$   &   $T_{2\textrm{E}}^\text{avg}$   &   $T_{2\textrm{E}}^\text{max}$    \\
                  & ($\mu$m)    &   Vacuum      &   (GHz)    &  ($\mu$s)   &   ($\mu$s)    &   ($\times 10^{6}$)   &   ($\times 10^{6}$)    &  ($\mu$s)   &   ($\mu$s)\\
       \hline
        1      &   70  &   HV    &   2.985   &   610.4 $\pm$ 14.5  &   1030.9 $\pm$ 34.1 &   11.4 $\pm$ 0.3 &   19.3 $\pm$ 0.6  &   297.5 $\pm$ 4.8 &   492.8 $\pm$ 122.4 \\
        2      &   70  &   HV    &   3.408   &   423.5 $\pm$ 9.6   &   762.6 $\pm$ 17.5  &   9.1 $\pm$ 0.2  &   16.3 $\pm$ 0.4  &   164.7 $\pm$ 2.1 &   228.4 $\pm$ 15.3 \\
        3      &   70  &   HV    &   3.398   &   579.8 $\pm$ 6.5   &   875.9 $\pm$ 49.4  &   12.4 $\pm$ 0.1 &   18.7 $\pm$ 1.1  &   128.6 $\pm$ 1.7 &   155.7 $\pm$ 5.0 \\
        4      &   70  &   HV    &   3.371   &   359.1 $\pm$ 4.9   &   464.6 $\pm$ 9.9   &   7.6 $\pm$ 0.1  &   9.8 $\pm$ 0.2   &   109.1 $\pm$ 1.2 &   141.6 $\pm$ 4.9 \\
        5      &   70  &   HV    &   3.791   &   375.9 $\pm$ 5.7   &   633.4 $\pm$ 48.2  &   9.0 $\pm$ 0.1  &   15.1 $\pm$ 1.2  &   57.6 $\pm$ 0.4  &   67.8 $\pm$ 5.9 \\
        6      &   70  &   HV    &   3.122   &   441.4 $\pm$ 12.5  &   717.9 $\pm$ 36.7  &   8.7 $\pm$ 0.2  &   14.1 $\pm$ 0.7  &   --   &   -- \\
        7      &   70  &   HV    &   3.108   &   570.1 $\pm$ 3.8   &   875.5 $\pm$ 18.2  &   11.1 $\pm$ 0.1 &   17.1 $\pm$ 0.4  &   188.4 $\pm$ 3.7 &   351.8 $\pm$ 50.2 \\
        8      &   70  &   HV    &   3.437   &   343.8 $\pm$ 7.4   &   534.5 $\pm$ 13.7  &   7.4 $\pm$ 0.2  &   11.5 $\pm$ 0.3  &   --   &   -- \\
        9      &   70  &   HV    &   3.823   &   450.7 $\pm$ 7.7   &   700.6 $\pm$ 26.6  &   10.8 $\pm$ 0.2 &   16.8 $\pm$ 0.6  &   --   &   -- \\
        10     &   70  &   HV    &   3.436   &   522.5 $\pm$ 11.3  &   710.6 $\pm$ 47.8  &   11.3 $\pm$ 0.2 &   15.3 $\pm$ 1.0  &   --   &   -- \\
        11     &   70  &  HV   &   3.702   &   287.7 $\pm$ 3.7   &   518.1 $\pm$ 41.3  &   6.7 $\pm$ 0.1  &   12.1 $\pm$ 1.0    &   --   &   -- \\
        12     &   70  &   HV    &   3.776   &   343.2 $\pm$ 5.5   &   749.9 $\pm$ 33.8  &   8.1 $\pm$ 0.1  &   17.8 $\pm$ 0.8  &   --   &   -- \\
        13     &   70  &   HV    &   3.947   &   588.5 $\pm$ 3.0   &   799.9 $\pm$ 45.5  &   14.6 $\pm$ 0.1 &   19.8 $\pm$ 1.1  &   --   &   --  \\
        14     &   70  &   HV    &   4.059   &   253.4 $\pm$ 2.3   &   403.1 $\pm$ 43.0  &   6.5 $\pm$ 0.1  &   10.3 $\pm$ 1.1  &   --   &   --  \\
        15     &   70  &   HV    &   4.413   &   321.8 $\pm$ 3.4   &   603.7 $\pm$ 38.7  &   8.9 $\pm$ 0.1 &   16.7 $\pm$ 1.1   &   --   &   --  \\
        16     &   70  &   HV    &   3.010   &   794.2 $\pm$ 8.7   &   1265.6 $\pm$ 34.5 &   15.0 $\pm$ 0.2 &   23.9 $\pm$ 0.7  &   331.2 $\pm$ 3.7 &   491.8 $\pm$ 37.7 \\
        17     &   100 &   HV    &   3.316   &   521.9 $\pm$ 8.5   &   1015.5 $\pm$ 39.1 &   10.9 $\pm$ 0.2 &   21.2 $\pm$ 0.8  &   695.2 $\pm$ 13.4    &   1184.5 $\pm$ 98.2 \\
        18     &   100 &   HV    &   3.391   &   623.1 $\pm$ 9.0   &   1071.4 $\pm$ 23.1 &   13.3 $\pm$ 0.2 &   22.8 $\pm$ 0.5  &   --   &   --  \\
        19     &   80  &   HV    &   3.701   &   417.3 $\pm$ 9.3   &   833.5 $\pm$ 30.1  &   9.7 $\pm$ 0.2 &   19.4 $\pm$ 0.7   &   --   &   --  \\
        20     &   70  &   HV    &   4.020   &   396.7 $\pm$ 5.6   &   606.8 $\pm$ 11.7  &   10.0 $\pm$ 0.1 &   15.3 $\pm$ 0.3  &   --   &   --  \\
        21    &   70  &   HV  &   3.863   &   481.7 $\pm$ 4.1   &   730.8 $\pm$ 24.3  &   11.7 $\pm$ 0.1 &   17.7 $\pm$ 0.6  &   --   &   --   \\
        22    &   70  &   HV  &   4.255   &   375.1 $\pm$ 3.2   &   538.0 $\pm$ 10.5  &   10.0 $\pm$ 0.1 &   14.4 $\pm$ 0.3  &   --   &   --  \\
        23    &   70  &   HV    &   4.965   &   240.7 $\pm$ 2.8   &   415.7 $\pm$ 13.8  &   7.51 $\pm$ 0.1 &   13.0 $\pm$ 0.4  &   --   &   --  \\
        24     &   70  &   HV    &   2.943   &   533.5 $\pm$ 7.7   &   854.8 $\pm$ 55.2  &   9.9 $\pm$ 0.1 &   15.8 $\pm$ 1.0  &   --   &   --  \\
        25     &   70  &   HV    &   3.312   &   382.8 $\pm$ 7.1   &   703.4 $\pm$ 61.2  &   8.0 $\pm$ 0.1 &   14.6 $\pm$ 1.3  &   --   &   --  \\
        26     &   70  &   HV    &   3.192   &   387.3 $\pm$ 6.7   &   715.5 $\pm$ 75.0  &   7.8 $\pm$ 0.1 &   14.3 $\pm$ 1.5  &   --   &   --  \\
        27     &   70  &   HV    &   3.656   &   241.3 $\pm$ 1.4   &   306.7 $\pm$ 13.3  &   5.5 $\pm$ 0.1 &   7.0 $\pm$ 0.3  &   --   &   --  \\
        28     &   70  &   HV    &   3.660   &   421.6 $\pm$ 4.9   &   621.9 $\pm$ 44.7  &   9.7 $\pm$ 0.1 &   14.3 $\pm$ 1.0  &   --   &   --  \\
        29     &   70  &   HV    &   3.496   &   363.3 $\pm$ 4.4   &   553.5 $\pm$ 65.5  &   8.0 $\pm$ 0.1 &   12.2 $\pm$ 1.4  &   --   &   --  \\
        30     &   70  &   UHV    &   2.741   &   646.7 $\pm$ 9.2   &   1234.8 $\pm$ 133.6 &   11.1 $\pm$ 0.2 &   21.3 $\pm$ 2.3  &   --   &   --  \\
        31     &   70  &   UHV    &   3.204   &   516.1 $\pm$ 8.8   &   1013.3 $\pm$ 47.0  &   10.4 $\pm$ 0.2 &   20.4 $\pm$ 1.0    &   416.6 $\pm$ 6.2 &   635.3 $\pm$ 52.0 \\
        32     &   70  &   UHV    &   3.571   &   489.1 $\pm$ 4.8   &   946.3 $\pm$ 90.8  &   11.0 $\pm$ 0.1 &   21.2 $\pm$ 2.0 &   446.3 $\pm$ 3.3 &   601.7 $\pm$ 52.4 \\
        33     &   70  &   UHV    &   3.908   &   397.9 $\pm$ 3.2   &   654.0 $\pm$ 21.3   &   9.8 $\pm$ 0.1 &   16.1 $\pm$ 0.5 &   269.1 $\pm$ 2.4 &   393.4 $\pm$ 29.9 \\
        34     &   70  &   UHV    &   2.407   &   1003.0 $\pm$ 13.4  &   1681.8 $\pm$ 100.9  &   15.2 $\pm$ 0.2 &   25.4 $\pm$ 1.5  &   --   &   --  \\
        35     &   70  &   UHV    &   2.741   &   464.4 $\pm$ 6.6    &   859.9 $\pm$ 109.6  &   8.0 $\pm$ 0.1 &   14.8 $\pm$ 1.9  &   --   &   --  \\
        36     &   70  &   UHV    &   3.109   &   511.6 $\pm$ 3.5   &   735.3 $\pm$ 60.3  &   10.0 $\pm$ 0.1 &   14.4 $\pm$ 1.2 &   693.1 $\pm$ 16.0    &   1058.2 $\pm$ 82.0 \\
        37     &   70  &   UHV    &   3.440   &   278.8 $\pm$ 4.4   &   451.7 $\pm$ 30.5  &   6.0 $\pm$ 0.1 &   9.8 $\pm$ 0.1  &   --   &   --  \\
        38     &   70  &   UHV    &   3.451   &   481.4 $\pm$ 6.6   &   728.5 $\pm$ 62.0  &   10.4 $\pm$ 0.1 &   15.8 $\pm$ 1.3  &   --   &   --  \\
        39     &   70  &   UHV    &   3.829   &   378.3 $\pm$ 5.5  &   650.8 $\pm$ 125.6  &   9.1 $\pm$ 0.1 &   15.7 $\pm$ 3.0  &   744.3 $\pm$ 16.2    &   955.2 $\pm$ 63.0 \\
        40     &   70  &   UHV    &   3.074   &   520.0 $\pm$ 6.0   &   831.6 $\pm$ 82.7 &   10.0 $\pm$ 0.1 &   16.1 $\pm$ 1.6  &   686.3 $\pm$ 9.8 &   1048.9 $\pm$ 93.2 \\
        41     &   70  &   UHV    &   3.227   &   538.7 $\pm$ 3.9   &   772.3 $\pm$ 41.1 &   10.9 $\pm$ 0.1 &   15.7 $\pm$ 0.1  &   647.9 $\pm$ 7.8 &   939.9 $\pm$ 126.4 \\
        42     &   70  &   UHV    &   3.418   &   500.6 $\pm$ 4.9   &  752.1 $\pm$ 38.9 &   10.8 $\pm$ 0.1 &   16.2 $\pm$ 0.1   &   676.5 $\pm$ 9.7 &   953.1 $\pm$ 91.9 \\
        43     &   70  &   UHV    &   3.381   &   415.8 $\pm$ 6.4   &   750.7 $\pm$ 82.3  &   8.8 $\pm$ 0.1 &   15.9 $\pm$ 1.8  &   354.4 $\pm$ 6.8 &   579.6 $\pm$ 54.3 \\
        44     &   70  &   UHV    &   3.634   &   371.2 $\pm$ 3.4   &   609.4 $\pm$ 31.9  &   8.5 $\pm$ 0.1 &   13.9 $\pm$ 0.1  &   569.9 $\pm$ 7.6 &   795.1 $\pm$ 59.2 \\
        45     &   70  &   UHV    &   3.831   &   324.9 $\pm$ 2.7   &   471.6 $\pm$ 38.9  &   7.8 $\pm$ 0.1 &   11.4 $\pm$ 0.1  &   537.3 $\pm$ 5.7 &   773.8 $\pm$ 84.6  \\
   \end{tabular}
   \end{ruledtabular}
\end{table*}

\begin{table*}
\centering
\caption{\label{table:T1s_CL} Device parameters, lifetime, and coherence times for 12 Ta-on-Si qubits with drive lines fabricated across 2 chips.}
\begin{ruledtabular}
\begin{tabular}{cccccccccc}
        Qubit     & Pitch  &   Deposition   &   Frequency   &   $T_{1}^\text{avg}$   &   $T_{1}^\text{max}$   &   $Q^\text{avg}$   &   $Q^\text{max}$  &   $T_{2\textrm{E}}^\text{avg}$   &   $T_{2\textrm{E}}^\text{max}$  \\
                     & ($\mu$m)    &   Vacuum     &   (GHz)    &  ($\mu$s)   &   ($\mu$s)    &   ($\times 10^{6}$)   &   ($\times 10^{6}$)  &  ($\mu$s)   &   ($\mu$s)\\
       \hline
        1      &   70  &   UHV    &   3.274   &   562.7 $\pm$ 11.8  &   638.2 $\pm$ 18.0 &   11.6 $\pm$ 0.2 &   13.1 $\pm$ 0.4  &   285.1 $\pm$ 9.6 &   319.8 $\pm$ 29.6 \\
        2      &   70  &   UHV    &   3.337   &   559.5 $\pm$ 25.9   &   729.4 $\pm$ 13.1  &   11.7 $\pm$ 0.1 &   15.3 $\pm$ 0.4  &   212.8 $\pm$ 10.5 &   290.5 $\pm$ 17.6 \\
        3      &   70  &   UHV    &   3.500   &   549.7 $\pm$ 19.4   &   706.0 $\pm$ 16.5  &   12.1 $\pm$ 0.4 &   15.5 $\pm$ 0.3  &   208.0 $\pm$ 15.8 &   395.1 $\pm$ 55.4 \\
        4      &   70  &   UHV    &   3.652   &   440.0 $\pm$ 22.0   &   569.8 $\pm$ 11.1   &   10.1 $\pm$ 0.5  &   13.1 $\pm$ 0.3  &   178.6 $\pm$ 10.7 &   258.5 $\pm$ 19.0 \\
        5      &   70  &   UHV    &   3.756   &   291.9 $\pm$ 14.1   &   382.84 $\pm$ 16.8  &   6.89 $\pm$ 0.3  &   9.03 $\pm$ 0.4  &   - &   - \\
        6      &   70  &   UHV    &   3.942   &   354.0 $\pm$ 5.26  &   386.49 $\pm$ 7.14  &   8.77 $\pm$ 0.1  &   9.57 $\pm$ 0.2  &   158.9 $\pm$ 9.7 &   255.9 $\pm$ 21.9 \\
        7      &   70  &   UHV    &   3.215   &   523.7 $\pm$ 28.0   &   782.3 $\pm$ 20.1  &   10.6 $\pm$ 0.1 &   15.8 $\pm$ 0.4  &   86.5 $\pm$ 5.5 &   141.9 $\pm$ 14.1 \\
        8      &   70  &   UHV    &   3.337   &   290.4 $\pm$ 9.4   &   380.88 $\pm$ 28.6  &   6.1 $\pm$ 0.2  &   7.99 $\pm$ 0.6  &   72.6 $\pm$ 10.5 &   135.0 $\pm$ 24.6 \\
        9      &   70  &   UHV    &   3.423  &   332.76 $\pm$ 20.7   &   431.9 $\pm$ 18.7  &   7.16 $\pm$ 0.4 &   9.29 $\pm$ 0.4  &  196.1 $\pm$ 12.1 &   297.6 $\pm$ 38.5 \\
        10     &   70  &   UHV    &   3.595   &   341.3 $\pm$ 10.2  &   442.0 $\pm$ 10.4  &   7.71 $\pm$ 0.2 &   9.98 $\pm$ 0.2  &   371.4 $\pm$ 12.1 &   461.5 $\pm$ 46.2 \\
        11     &   70  &   UHV    &   3.747   &   410.9 $\pm$ 12.6   &   506.4 $\pm$ 19.2  &   9.7 $\pm$ 0.3  &   11.9 $\pm$ 0.5  &   298.6 $\pm$ 13.8 &   376.2 $\pm$ 46.6 \\
        12     &   70  &   UHV    &   3.872   &   81.8 $\pm$ 2.91   &   106.6 $\pm$ 5.26  &   1.99 $\pm$ 0.1  &   2.59 $\pm$ 0.1  &   91.5 $\pm$ 5.38 &   162.8 $\pm$ 29.6 
        \\
   \end{tabular}
   \end{ruledtabular}
\end{table*}

\begin{table}[h]
    \centering
    \begin{tabular}{|c|c|c|c|c|c|c|}
        \hline
        {$\omega_q/2\pi$~(GHz)} & {$\omega_r/2\pi$~(GHz)} & {$E_c/2\pi$~(MHz)} & {$\kappa/2\pi$~(MHz)} & {$\chi/2\pi$~(MHz)} & {$g/2\pi$~(MHz)} & {$T_{1,p} ~(ms)$} \\
        \hline
        $3-4$   & $6.8-7.3$ & $200$  & $0.2-0.4$   & $0.2-0.3$   & $50-60$   & $100$   \\
        \hline
    \end{tabular}
    \caption{Transmon device parameters}
    \label{table:devparams}
\end{table}

\end{document}